\documentclass[aps,prb,epsf,twocolumn,showpacs]{revtex4-2}

\usepackage{CJK,amsmath,amssymb,graphics,epsfig,epstopdf,color,verbatim,ulem,braket,tabularx,float,makecell,multirow}
\usepackage[colorlinks,linkcolor=blue,citecolor=blue,urlcolor=blue]{hyperref}

\begin{document}
\begin{CJK*}{GBK}{song}
\title{\mbox{Phase diagram and magnetic excitations of $J_1$--$J_3$ Heisenberg}
\mbox{model on the square lattice}}

\author{Muwei Wu$^{1}$}
\author{Shou-Shu Gong$^{2}$}
\email{shoushu.gong@buaa.edu.cn}
\author{Dao-Xin Yao$^{1,3}$}
\email{yaodaox@mail.sysu.edu.cn}
\author{Han-Qing Wu$^{1}$}
\email{wuhanq3@mail.sysu.edu.cn}
\affiliation{\mbox{$^{1}$Center for Neutron Science and Technology, Guangdong Provincial Key Laboratory of Magnetoelectric Physics and Devices,}
\mbox{State Key Laboratory of Optoelectronic Materials and Technologies,}
\mbox{School of Physics, Sun Yat-sen University, Guangzhou, 510275, China}
\mbox{$^{2}$Department of Physics and Peng Huanwu Collaborative Center for Research and Education,}
\mbox{Beihang University, Beijing 100191, China}\\
\mbox{$^{3}$International Quantum Academy, Shenzhen 518048, China}}

\begin{abstract}
  We study the phase diagram and the dynamical spin structure factor of the spin-1/2 $J_1$--$J_3$ Heisenberg model on the square lattice using density matrix renormalization group, exact diagonalization (ED), and cluster perturbation theory (CPT). By extrapolating the order parameters and studying the level crossings of the low-lying energy and entanglement spectra, we obtain the phase diagram of this model and identify a narrow region of quantum spin liquid (QSL) phase followed by a plaquette valence-bond solid  (PVBS) state in the intermediate region, whose nature has been controversial for many years. More importantly, we use CPT and ED to study the dynamical spin structure factor in the QSL and the PVBS phase. In the QSL phase, the high-energy magnon mode completely turns into some dispersive weak excitations around the $X$ and $M$ points. For the PVBS phase, the low-energy spectrum is characterized by a gapped triplet excitation, and at the high energy, we find another branch of dispersive excitation with broad continua, which is unlike the plaquette phase in the 2$\times$2 checkerboard model. In the latter case, the second branch of excitation is nearly flat due to the weak effective interactions between the local excitations of the plaquettes. And in the $J_1$--$J_3$ Heisenberg model, the uniform interactions and the spontaneously translational symmetry breaking of the PVBS phase make the difference in the excitation spectra.
\end{abstract}


\date{\today}
\maketitle
\end{CJK*}

\section{Introduction}

Geometry frustration and competing interactions in quantum magnets can lead to many novel phenomena~\cite{Lacroix2011}, in which the quantum spin liquid (QSL) states~\cite{XGWen1991, *Balents2010, *XGWen2002, *Kitaev2006, *Savary2016, *Norman2016, *ZYi2017, *Broholm2020} have gained much research interest in recent years.
A QSL is an exotic quantum state beyond the Landau-Ginzburg-Wilson (LGW) paradigm, in which the conventional magnetic order and dimer order are prevented from being developed even at zero temperature.
More interestingly, QSL states have long-range entanglement and emergent fractionalized excitations such as the neutral spinon, which obey the anyon statistics~\cite{XGWen1991, *Balents2010, *XGWen2002, *Kitaev2006, *Savary2016, *Norman2016, *ZYi2017, *Broholm2020}.
Therefore, while the absent conventional orders provide a necessary condition for detecting a QSL, the magnetic excitations that can be probed by inelastic neutron scattering experiment are playing a more important role in the identification of QSL states in both models and materials.

One of the most important models to search and study QSL is the square-lattice Heisenberg model with competing interactions.
Due to the close relation with the parent compound of high-temperature superconductors~\cite{Anderson1987, XGWen2006}, the $J_1$--$J_2$ square-lattice Heisenberg model has attracted a lot of studies on both the ground-state phase diagram and the excitation spectrum, by various methods such as exact diagonalization (ED)~\cite{Dagotto1989, Schulz1996, Capriotti2000, Mambrini2006, Richter2010}, density matrix renormalization group (DMRG)~\cite{HChJiang2012, SSGong2014, LWang2018}, tensor network (TN)~\cite{Murg2009, JFYu2012, LWang2013, LWang2016, Haghshenas2018, WYLiu2018, Didier2019, WYLiu2021tri2}, variational Monte Carlo (VMC)~\cite{Capriotti2001, TLi2012, Mezzacapo2012, WJHu2013, YQi2014, Morita2015, Ferrari2018, Ferrari2020}, and many other methods~\cite{Chandra1988, Gelfand1989, Read1991, Singh1999, GMZhang2003, Darradi2008, YRen2014, ShLYu2018, Nomura2021tri2, Shackleton2021}.
These studies have proposed different candidate states for the intermediate paramagnetic region.
Up to now, the consensus is that below the stripe magnetic phase, the system is in a weak valence-bond solid (VBS) phase, although its nature is still under debate between a plaquette VBS (PVBS) and a columnar VBS (CVBS).
Between the N\'eel and the VBS phases, some new results support a gapless QSL phase~\cite{LWang2018, WYLiu2021tri2, Morita2015, Ferrari2020, Nomura2021tri2}, but its nature is far from clear.
Meanwhile, due to the limit of system size in these studies, a direct N\'eel-VBS transition may not be excluded. For the magnetic excitation, a cluster perturbation theory (CPT) study~\cite{ShLYu2018} and a VMC study~\cite{Ferrari2018} both supported the existence of a gapless Z$_2$ spin liquid and found that the spectrum of this phase is characterized by a broad continuum. In experiment, there are many compound materials which may be effectively described by the $J_1$--$J_2$ Heisenberg model, such as Li$_2$VO$M$O$_4$ ($M$ = Si, Ge)~\cite{Melzi2000, Bombardi2004}, VOMoO$_4$~\cite{Carretta2002, Bombardi2005}, PbVO$_3$~\cite{Tsirlin2008}, and Sr$_2$CuTe$_{1-x}$W$_x$O$_6$~\cite{Babkevich2016, Koga2016, Mustonen2018tri1, Mustonen2018tri2, Katukuri2020, WShHong2021}.
QSL-like behaviors have been reported in some of these compounds~\cite{Mustonen2018tri1, Mustonen2018tri2, Katukuri2020, WShHong2021}.

To establish a better understanding of the competing phases in the square-lattice Heisenberg model, the $J_1$--$J_3$ model has also been extensively studied in the past three decades~\cite{Gelfand1989, Locher1990, Moreo1990, Chubukov1991tri2, Rastelli1992, Ferrer1993, Leung1996, Capriotti2004tri1, Capriotti2004tri2, Mambrini2006, Murg2009, Sindzingre2010, Reuther2011, Kharkov2018, WYLiu2021tri3, Daniel2021}.
In the classical limit, the model has a N\'eel antiferromagnetic (NAF) phase and a spiral order phase, which are separated at $J_3 / J_1$ = 0.25~\cite{Gelfand1989, Locher1990, Moreo1990, Chubukov1991tri2, Rastelli1992, Ferrer1993}.
After considering quantum fluctuations, a paramagnetic region also emerges near $J_3 / J_1$ = 0.5.
An early DMRG study~\cite{Capriotti2004tri1} on the ladder with open boundary conditions suggested that the dimer order would vanish with the increase of the leg number and the ground state near $J_3$ = 0.5 may be a QSL.
In contrast, an ED study~\cite{Mambrini2006} and a projected entangled pair state (PEPS) study~\cite{Murg2009} found that the ground state is likely to be a PVBS state, which was also supported by the series expansion (SE), coupled cluster method (CCM), and dynamical functional renormalization group (FRG) calculations~\cite{Reuther2011}.
A TN study further claimed that this PVBS state belongs to a higher-order symmetry-protected topological phase~\cite{Daniel2021}.
Very recently, a large-scale PEPS study~\cite{WYLiu2021tri3} showed that there is a remarkable QSL phase (0.28 $\lesssim J_3 / J_1 \lesssim$ 0.38) between the N\'eel and the VBS phases, which is inherited from the QSL in the $J_1$--$J_2$ model.
This phase diagram provides a fascinating perspective to understand the emergent QSL phase from the neighbor deconfined quantum critical point (DQCP)~\cite{Senthil2004tri1, Senthil2004tri2, Sandvik2007}. The NAF-QSL and QSL-VBS phase transitions appear to be continuous, and the obtained critical exponents suggest that the two transitions may belong to new types of universality classes~\cite{WYLiu2021tri3}.
However, the nature of the VBS state was not determined in this large-scale simulation and the existence of the QSL phase was only probed by static properties.
Therefore, in this paper, we reexamine the phase diagram of this model by combining the ED and DMRG calculations.
We also explore the magnetic excitations in the VBS and the possible QSL phase, which are compared with the results of the $J_1 - J_2$ model as well.

\begin{figure}[t]
  \centering
  \includegraphics[width=0.49\textwidth]{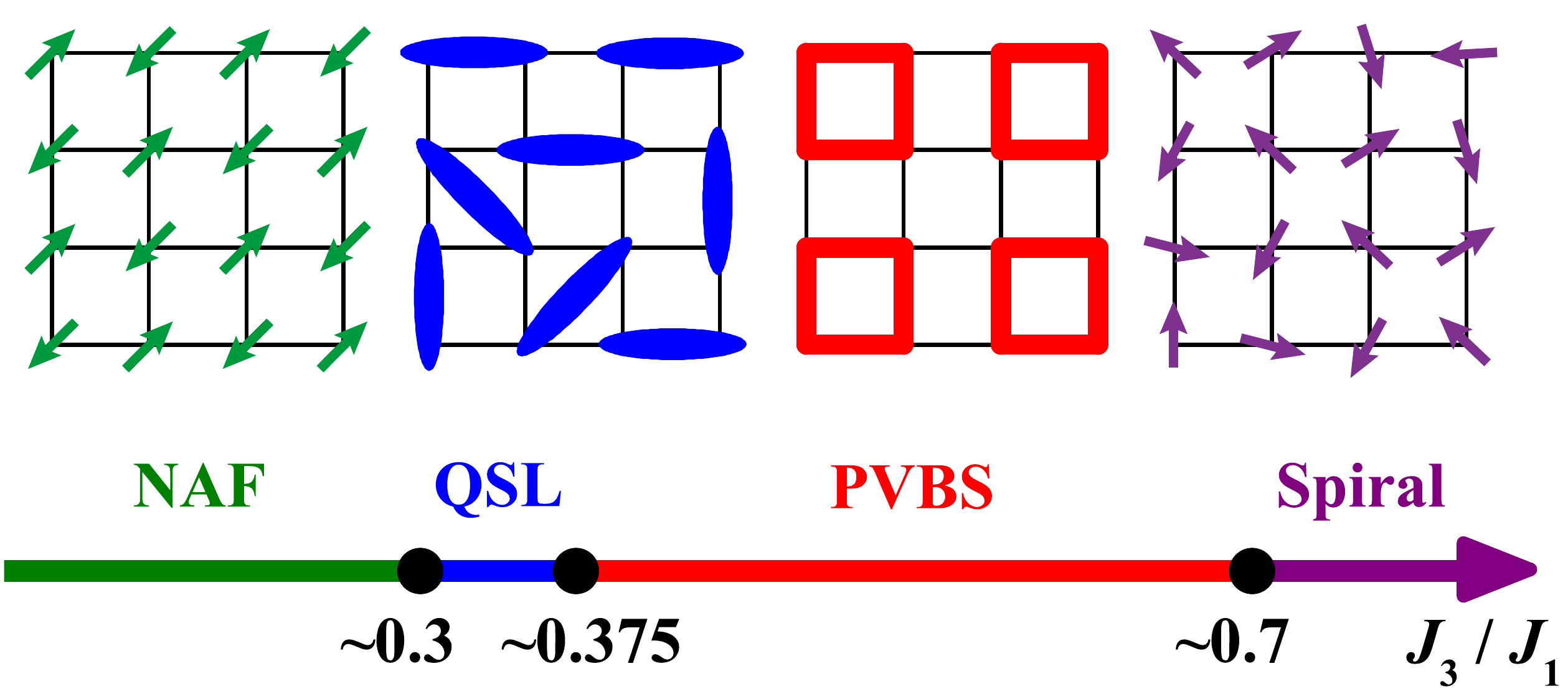}
  \caption{Phase diagram of the spin-$1/2$ $J_1$--$J_3$ Heisenberg model on the square lattice. When $J_3$ is small, the model is in the N\'eel antiferromagnetic (NAF) phase. With growing $J_3$, it enters into a quantum spin liquid (QSL) phase for 0.3 $\lesssim J_3/J_1 \lesssim 0.375$. In larger $J_3$, there is a region of plaquette valence bond solid (PVBS) phase. Around $J_3/J_1 \simeq$ 0.7, a phase transition between the PVBS phase and the spiral magnetic order phase occurs.}
  \label{fig:PhaseDiagram}
\end{figure}

The rest of this paper is organized as follows. In Sec.~\ref{Sec:Model}, we introduce the model Hamiltonian and define the order parameters as well as the physical observables that we will calculate. In Sec.~\ref{Sec:PhaseDiagram}, we show the phase diagram obtained by our DMRG and ED results. As shown in Fig.~\ref{fig:PhaseDiagram}, we identify that the paramagnetic region actually includes two phases: a QSL phase and a PVBS phase based on the calculation of the order parameters, energy spectrum, and entanglement spectrum. More importantly, in Sec.~\ref{Sec:Excitation}, we show the dynamical spin structure factor for different phases by CPT and ED. In the QSL phase, we find signals of some weak continua around the $M$ and $X$ points. And our study of the magnetic excitation in the PVBS phase will provide more theoretical understanding of the PVBS phase with spontaneously translational symmetry breaking. Finally, we provide a summary and discussion in Sec.~\ref{Sec:Summary}.

\section{Model and Method}
\label{Sec:Model}

The Hamiltonian of the spin-$1/2$ $J_1$--$J_3$ Heisenberg model on the square lattice reads
\begin{equation*}
\begin{split}\begin{array}{l}
H = J_1 \sum\limits_{\left\langle {i,j} \right\rangle} {\hat{\mathbf{S}}_{i} \cdot \hat{\mathbf{S}}_{j}} + J_3 \sum\limits_{\left\langle\left\langle\left\langle {i,j} \right\rangle\right\rangle\right\rangle} {\hat{\mathbf{S}}_{i} \cdot \hat{\mathbf{S}}_{j}}.
\end{array}
\label{Eq:Hmlt}
\end{split}
\end{equation*}
In the following calculation, we set $J_1 = 1$ as the energy unit.
We use SU(2) DMRG~\cite{SRWhite1992, McCulloch2002} and ED to study the ground-state phase diagram.
For the magnetic excitation, we employ the ED and CPT~\cite{Gros1993, Senechal2000, ShLYu2018} methods.
The finite-size clusters with the periodic boundary conditions (torus geometry) used in the ED calculation are shown in Appendix~\ref{App:FSClusters}.
The cylinder geometry we used in the DMRG calculation is the rectangular cylinder with the periodic boundary conditions in the $y$ direction and the open boundary conditions in the $x$ direction.
To reduce the boundary effects, we choose the lattice size with $L_x \geq 2 L_y$ ($L_x$ and $L_y$ represent the numbers of sites in the $x$ and $y$ directions, respectively) and take the correlations of the middle $L_y \times L_y$ sites to calculate the order parameters~\cite{SRWhite2007, HChJiang2012, SSGong2014}.

\begin{figure}[t]
  \centering
  \includegraphics[width=0.49\textwidth]{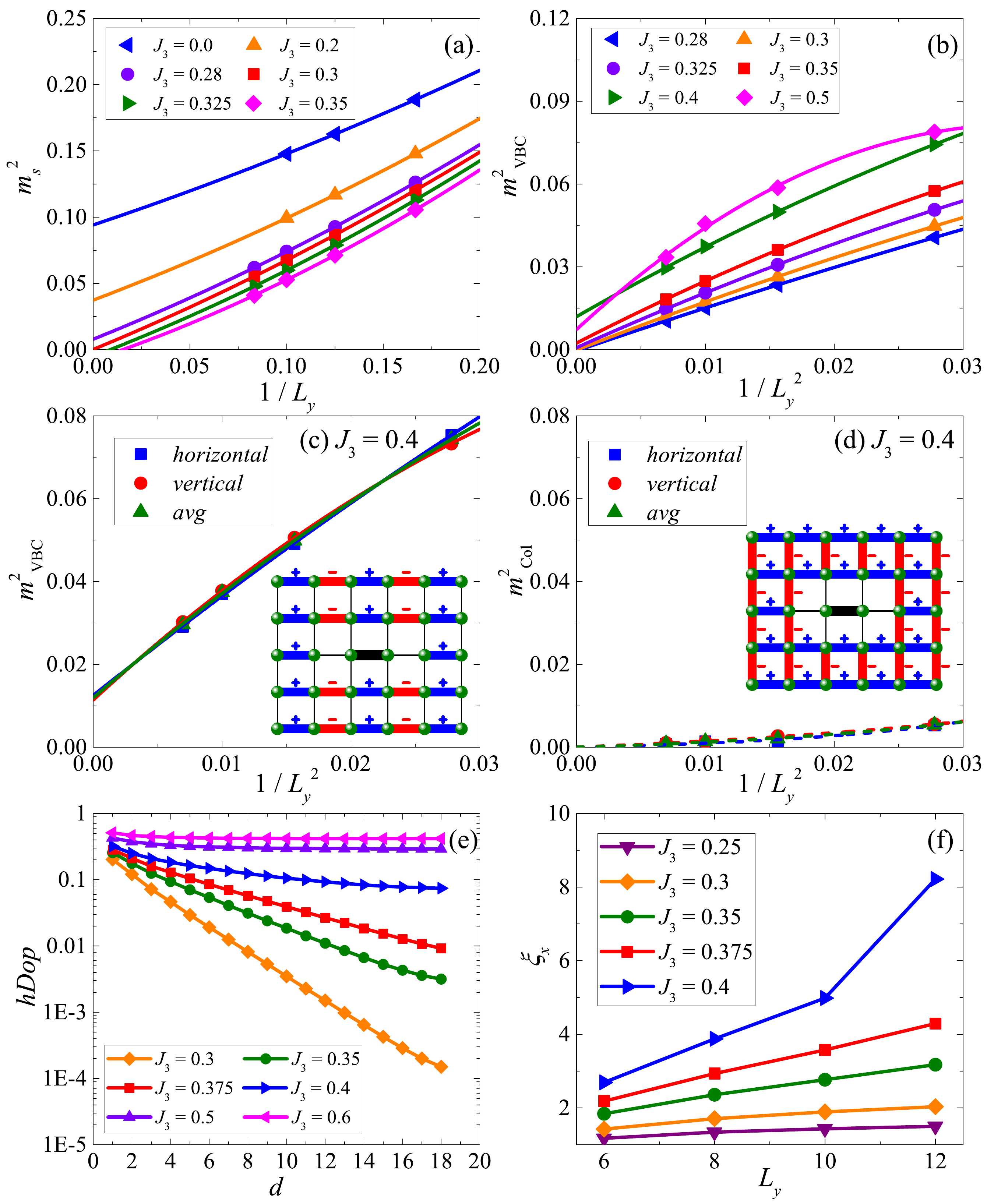}
  \caption{Finite-size extrapolation of (a) the N\'eel order parameter $m_{s}^{2}$ and (b) the dimer order parameter $m_{\textrm{VBC}}^{2}$. (c) and (d) are the extrapolation of the dimer order parameters $m_{\textrm{VBC}}^{2}$ and $m_{\textrm{Col}}^{2}$ at $J_3$ = 0.4 obtained with a horizontal reference bond, a vertical reference bond, and the averaged results of these two kinds of reference bonds. The lines in (a)--(c) are the second-order polynomial fits for the data. And the dashed lines in (d) are guides to the eye. The insets in (c) and (d) show the phase factors $\epsilon_{\lambda}$ for $m_{\textrm{VBC}}^{2}$ and $m_{\textrm{Col}}^{2}$, respectively~\cite{Mambrini2006}. The bold bonds in the center of the insets represent the reference bond $(i, j)$ in the calculation of the dimer order parameters $m_{\lambda}^{2}$. (e) The logarithmic-linear plot of the hDOP with different $J_3$. $d$ is the distance measured from the boundary. The fitting of the exponential decay of the hDOP gives the decay length $\xi_x$. The hDOP at $J_3$ = 0.5, 0.6 are calculated on the cylinder geometry with $L_y$ = 10, and the other data are obtained on the cylinder geometry with $L_y$ = 12. (f) The decay length $\xi_x$ of the horizontal dimer order vs $L_y$ obtained by DMRG on the cylinder geometry.}
  \label{fig:MagneticOrders}
\end{figure}

In order to determine the phase boundaries and detect possible dimer order in the intermediate paramagnetic phase, we calculate three kinds of order parameters.
The first one is the N\'eel order parameter,
\begin{equation*}
m_{s}^{2} = \frac{1}{N} S(\mathbf{M}), \quad S(\mathbf{q}) = \frac{1}{N}\sum_{i,j} \left\langle \hat{\mathbf{S}}_{i} \cdot \hat{\mathbf{S}}_{j} \right\rangle e^{i \vec{q} \cdot \left( \vec{r}_i - \vec{r}_j \right)},
\end{equation*}
where $\mathbf{M} = \left(\pi, \pi\right)$.
The other two kinds of order parameters are the dimer order parameters~\cite{Mambrini2006},
\begin{equation*}
m_{\lambda}^{2} = \frac{1}{N_b} S_{\lambda}, \quad
S_{\lambda} = \sum_{\left(k,l\right)}\epsilon_{\lambda} \left(k,l\right) C_{ijkl},
\end{equation*}
where $\lambda$ is either ``VBC" (valence bond crystal) or ``Col" (columnar), and $N_b$ is the number of bonds used in the calculation of $m_{\lambda}^{2}$.
These two dimer order parameters can be analyzed together to distinguish the CVBS and the $s$-wave PVBS state whose symmetry-breaking state adiabatically connects to the direct product of $|\varphi_0 \rangle$ shown in Fig.~\ref{fig:App_Plaquette}.
For both VBS states, $m_{\textrm{VBC}}^{2}$ would be finite in the thermodynamic limit, while $m_{\textrm{Col}}^{2}$ would be finite only for the CVBS state and would be zero for the $s$-wave PVBS state~\cite{Mambrini2006}.
The phase factors $\epsilon_{\lambda}$ of these two dimer order parameters are chosen as shown in the insets of Figs.~\ref{fig:MagneticOrders}(c) and \ref{fig:MagneticOrders}(d), where the reference bond $(i,j)$ locates in the middle of the cylinders, either along the $x$ (horizontal) direction or the $y$ (vertical) direction.
The dimer-dimer correlation function $C_{ijkl}$ is defined as
\begin{equation*}
C_{ijkl} = 4\left[ \left\langle \left( \hat{\mathbf{S}}_{i} \cdot \hat{\mathbf{S}}_{j} \right) \left( \hat{\mathbf{S}}_{k} \cdot \hat{\mathbf{S}}_{l} \right) \right\rangle - \left\langle \hat{\mathbf{S}}_{i} \cdot \hat{\mathbf{S}}_{j} \right\rangle \left\langle \hat{\mathbf{S}}_{k} \cdot \hat{\mathbf{S}}_{l} \right\rangle \right].
\end{equation*}
We have also calculated the dimer structure factor
\begin{equation*}
D(\mathbf{q}) = \frac{1}{2N}\sum\limits_{\left\langle i, j\right\rangle}\sum\limits_{\left\langle k, l\right\rangle}e^{-i\mathbf{q} \cdot \mathbf{r}_{ij,kl}}C_{ijkl},
\end{equation*}
where $\mathbf{r}_{ij,kl}$ means the displacement between the centers of two bonds which connect two pairs of nearest-neighbor sites, $\left\langle i, j\right\rangle$ and $\left\langle k, l\right\rangle$.

To study the magnetic excitation, we calculate the dynamical spin structure factor defined as
\begin{equation*}
\begin{split}
S^{+-}(\mathbf{q},\omega)&=\sum_{n}\left\{|\langle \psi_n | \hat{S}^+_{\bf q} | \psi_0 \rangle|^2 \delta\left[\omega-(E_n-E_0)\right]\right\},
\end{split}
\end{equation*}
where $\hat{S}_{\mathbf{q}}^+ = \frac{1}{N} \sum_i{e^{-i\mathbf{q} \cdot \mathbf{r_i}} \hat{S}_{i}^+}$ is the Fourier transform of the spin operator $\hat{S}_{i}^+$, and $| \psi_n \rangle$ is the eigenstate of the Hamiltonian with energy $E_n$.
Using ED, we calculate the dynamical spin structure factor $S^{+-}(\mathbf{q},\omega)$ on the 36-site cluster under the periodic boundary conditions. Although the 36-site cluster may still be small to get the correct estimation of the thermodynamic-limit results, the results obtained by ED have no approximation and can also capture some important characteristics of the different phases in the $J_1$--$J_3$ Heisenberg model, which will be shown in Sec.~\ref{Sec:Excitation}. And a small Lorentz broadening factor $\eta$ = 0.05 is used in order to observe the prominent excitation peaks more clearly.

We also calculate the dynamical spin structure factor by using the bosonic version of the CPT method, which has been successfully applied to the $J_1$--$J_2$ Heisenberg model on the square lattice~\cite{ShLYu2018}.
By using the ED to do exact calculation within a cluster under the open boundary condition and treating the intercluster interaction as perturbations, the CPT method can get the dynamical spin structure factor $S^{+-}(\mathbf{q},\omega)$ in the thermodynamic limit (more details can be found in Ref.~\cite{ShLYu2018}).
In this paper, we use the 6$\times$4 cluster and set $\eta$ = 0.15 in the CPT calculation.

\section{Numerical Results}

\subsection{Phase Diagram}
\label{Sec:PhaseDiagram}

First of all, we consider the N\'eel order parameter $m_{s}^{2}$, which is shown in Fig.~\ref{fig:MagneticOrders}(a).
By keeping up to 8000 SU(2) states to obtain convergent data in the DMRG calculation (the DMRG truncation errors are smaller than $1\times 10^{-5}$), we calculate $m_{s}^{2}$ on cylinders with $L_y$ = 6 - 12.
We exclude $L_y$ = 4 because the $J_3$ interaction along the $y$ direction would be doubly counted.
The second-order polynomial fitting of $m_{s}^{2}$ suggests that the N\'eel order vanishes at $J_3 \simeq 0.3$, instead of $J_3$ = 0.25 in the classical limit~\cite{Gelfand1989, Locher1990, Moreo1990, Chubukov1991tri2, Rastelli1992, Ferrer1993}.
The smooth vanishing of the extrapolated $m_{s}^{2}$ suggests a continuous phase transition.

Next, we use DMRG to calculate the dimer order parameters $m_{\textrm{VBC}}^{2}$ and $m_{\textrm{Col}}^{2}$. 
Compared with the N\'eel order parameter $m_{s}^{2}$, the dimer order parameters for large $J_3$ and large $L_y$ are more difficult to converge in the DMRG calculation, especially for $J_3 \gtrsim $ 0.4 and $L_y$ = 12.
Therefore, we extrapolate $m_{\textrm{VBC}}^{2}$ and $m_{\textrm{Col}}^{2}$ versus $1/m$ [$m$ is the number of the SU(2) states kept in DMRG calculation], which is shown in Fig.~\ref{fig:App_Fit}(a).
In Figs.~\ref{fig:MagneticOrders}(c) and \ref{fig:MagneticOrders}(d), we show $m_{\textrm{VBC}}^{2}$ and $m_{\textrm{Col}}^{2}$ at $J_3$ = 0.4, which are obtained by choosing different reference bonds.
The results obtained with a horizontal and a vertical reference bond are almost the same at $J_3$ = 0.4.
By a second-order polynomial fitting, $m_{\textrm{VBC}}^{2}$ is extrapolated to finite, while the values of $m_{\textrm{Col}}^{2}$ are always very small and approach zero in the thermodynamic limit.
For a CVBS state, both $m_{\textrm{VBC}}^{2}$ and $m_{\textrm{Col}}^{2}$ would be finite, but for an $s$-wave PVBS state, only $m_{\textrm{VBC}}^{2}$ would be finite~\cite{Mambrini2006}.
Therefore, our results indicate that the ground state at $J_3$ = 0.4 would be an $s$-wave PVBS, although some other kinds of VBS states still cannot be fully ruled out~\cite{Ralko2008, Sindzingre2010, Takahashi2020}. And the proposal of the PVBS state is further confirmed by the clear plaquette pattern of the nearest-neighbor bond energy obtained on the 8$\times$8 cluster with the fully open boundary conditions, which are shown in Figs.~\ref{fig:APP_BondEnergy}(b)--\ref{fig:APP_BondEnergy}(d).
To determine the phase region of this $s$-wave PVBS, we calculate $m_{\textrm{VBC}}^{2}$ with different reference bonds, and the finite-size extrapolation of the averaged $m_{\textrm{VBC}}^{2}$ at different $J_3$ is shown in Fig.~\ref{fig:MagneticOrders}(b).
For $J_3 \lesssim$ 0.35, the extrapolated $m_{\textrm{VBC}}^{2}$ is nearly zero, and it is clearly nonzero at $J_3$ = 0.4. With further increasing $J_3$ in the PVBS phase, as shown in Fig.~\ref{fig:APP_BondEnergy_Cylinder}(b) for $J_3 = 0.5$, the nearest-neighbor bond energies obtained on the cylinder geometry have strongly broken the translational symmetry and are highly dimerized in the $x$ direction, which indicates a strong static dimer order. In this case, the dimer-dimer correlation function after subtracting the bond energy as the background may fail to identify the dimer order. Due to this reason, the scaling behavior of the dimer order parameter $m_{\textrm{VBC}}^{2}$ at $J_3$ = 0.5 is different from that at smaller $J_3$, as shown in Fig.~\ref{fig:MagneticOrders}(b). And as shown in Fig.~\ref{fig:APP_BondEnergy_Cylinder}(a), at smaller $J_3$ such as $J_3$ = 0.4, the boundary-induced dimer order decays fast from the boundary to the bulk, and the bond energy is relatively uniform in the bulk of the cylinder. So, the dimer order parameter $m_{\textrm{VBC}}^{2}$ can successfully detect the dimer order at smaller $J_3$.

In order to determine the critical $J_3$ where the PVBS order develops more accurately, we also study the decay length of the horizontal dimer order parameter (hDOP)~\cite{HChJiang2012, Sandvik2012,  SSGong2014} on the cylinders with $L_x \geq$ 3 $L_y$.
The hDOP is defined as the difference between the bond energies of two adjacent horizontal nearest-neighbor bonds, which is easier to converge in the DMRG calculation compared with the dimer correlation function and the calculated dimer order parameters.
The hDOPs for different $J_3$ with $L_y \geq$ 10 are shown in Fig.~\ref{fig:MagneticOrders}(e).
For $J_3 \leq$ 0.375, the hDOPs are nearly linear curves in the logarithmic-linear plot, meaning that the hDOPs decay exponentially with distance $d$ from the boundary to the bulk.
Then we use a function hDOP $\sim e^{-d/\xi_x}$ to fit the data and get the decay lengths $\xi_x$ of the horizontal dimer order parameter, which are shown in Fig.~\ref{fig:MagneticOrders}(f).
For $J_3 \lesssim$ 0.375, $\xi_x$ grows slowly and seems to be finite in the large-size limit.
For $J_3 \gtrsim$ 0.375, $\xi_x$ grows faster than linear, especially for $L_y \geq$ 10, indicating the existence of the dimer order.
At $J_3 \simeq$ 0.375, $\xi_x$ increases almost linearly with $L_y$.
Therefore, we take $J_3 \simeq$ 0.375 as the critical $J_3$ where the system enters the PVBS phase, which is very close to $J_3$ = 0.38 obtained by the PEPS simulation~\cite{WYLiu2021tri3}.
With further growing $J_3$, one can see that even on the cylinder with $L_y$ = 10, the hDOP at $J_3$ = 0.5 and 0.6 are almost flat with distance $d$, which indicates the tendency of an increasing dimer order at $J_3$ = 0.5 and 0.6.

\begin{figure}[t]
  \centering
  \includegraphics[width=0.49\textwidth]{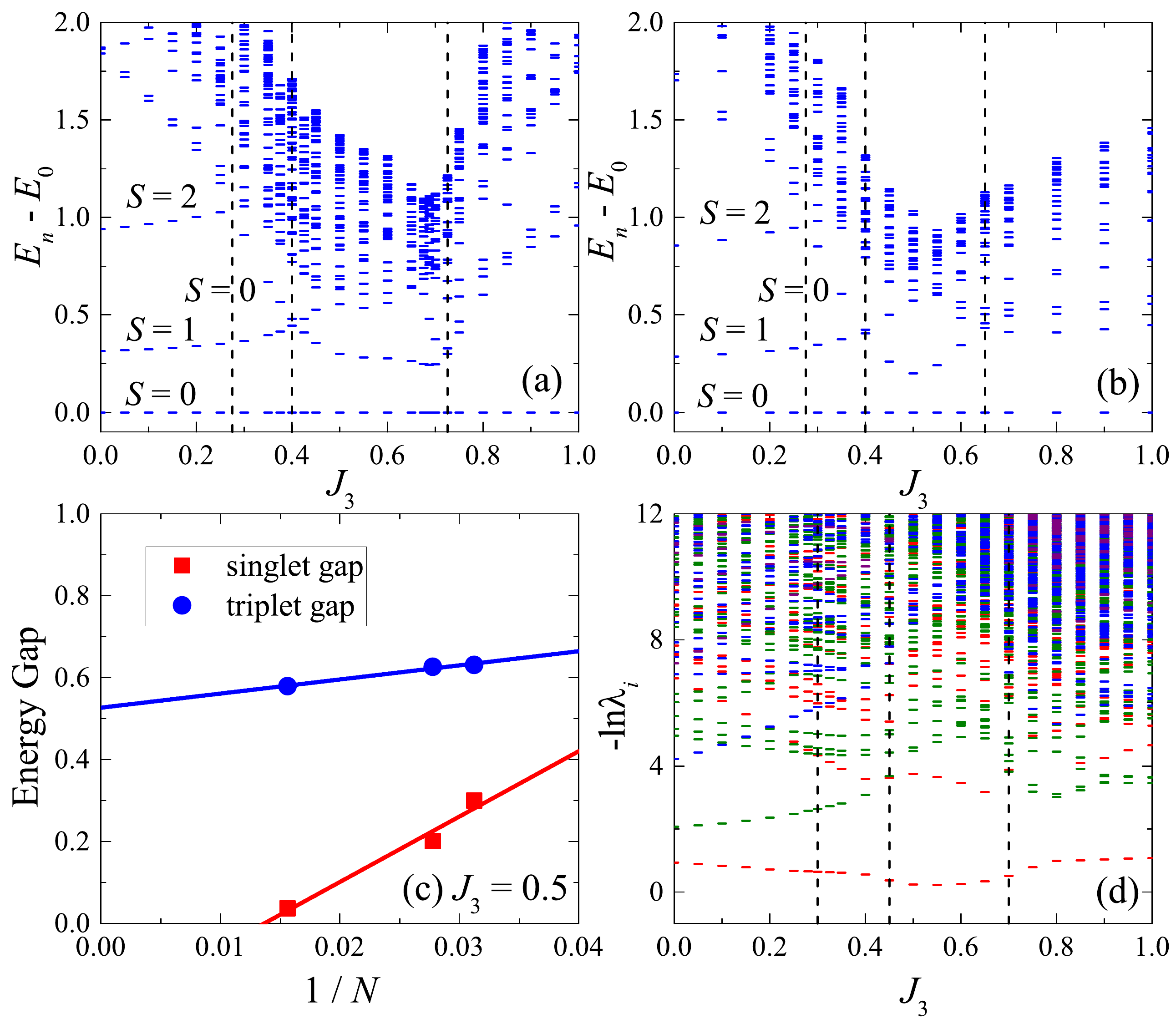}
  \caption{The low-energy spectra obtained on the (a) 32-site and (b) 36-site clusters. (c) The linear extrapolation of the singlet and triplet gap obtained on the torus geometry vs 1/$N$ at $J_3$ = 0.5. (d) The entanglement spectrum obtained on the cylinder geometry with $L_y = 8, L_x = 24$ by DMRG, in which the eigenvalues of the reduced density matrix $\lambda_i$ with different total spin $S = 0, 1, 2$ and $S > 2$ are represented by red, green, blue, and purple bars, respectively.}
  \label{fig:Spectrum}
\end{figure}

\begin{table*}[t]
  \caption{ The NAF-QSL, QSL-PVBS, and PVBS-Spiral transition points obtained by different physical observables. The corresponding methods and lattice sizes used to obtain these transition points are also listed.}
  \renewcommand{\arraystretch}{1.7} 
  \setlength\tabcolsep{15pt}
  \centering

  \begin{tabularx}{0.95\textwidth}{ c c c c c }
  \hline\hline
  Type & Physical Observable & Lattice Size & Method & $J_{3,c}$\ (error) \\
  \hline

  & $m^2_s$ & Cylinders ($L_y$ = 6 - 12) & DMRG & 0.3 (0.025) \\

  NAF-QSL & Energy & Torus (32\ sites) & ED & 0.275 (0.025)\\

  & Energy & Torus (36\ sites) & ED & 0.275 (0.025)\\

  & Entanglement & Cylinders ($L_y$ = 8) & DMRG & 0.3 (0.025)\\

  QSL-PVBS & $m^2_{VBC}$ & Cylinders ($L_y$ = 6 - 12) & DMRG & 0.35 (0.05) \\

  & $\xi_x$ & Cylinders ($L_y$ = 6 - 12) & DMRG & 0.375 (0.025) \\

  & Energy & Torus (32\ sites) & ED & 0.4 (0.025)\\

  & Energy & Torus (36\ sites) & ED & 0.4 (0.05)\\

  & Entanglement & Cylinders ($L_y$ = 8) & DMRG & 0.45 (0.05)\\

  PVBS-Spiral & Energy &Torus (32\ sites) & ED & 0.725 (0.025)\\

  & Energy & Torus (36\ sites) & ED & 0.65 (0.05)\\

  & Entanglement & Cylinders ($L_y$ = 8) & DMRG & 0.7 (0.05)\\
  \hline\hline

  \end{tabularx}
  \label{table:App_TransitionPoint}
\end{table*}

Besides order parameters, energy level crossing in the low-energy spectrum may also help to determine phase boundaries, which usually has a smaller finite-size effect than other physical observables and has been applied to the $J$--$Q$ model~\cite{Suwa2016} and the $J_1$--$J_2$ Heisenberg model on the square lattice~\cite{LWang2018, Nomura2021tri2}.
We also calculate the energy spectra for the $J_1$--$J_3$ Heisenberg model with 32 and 36 sites, which are shown in Figs.~\ref{fig:Spectrum}(a) and \ref{fig:Spectrum}(b), respectively.
When $J_3$ is small, the ground state is the N\'eel state and the lowest-energy states in the sectors with different total spin $S$ form the Anderson tower of states (TOS)~\cite{Anderson1952, Claire2005, Wietek2017}.
In the thermodynamic limit, these states are degenerate and the ground state will spontaneously break spin SU(2) symmetry; otherwise, on a finite-size lattice, they are nondegenerate and the finite-size excitation gaps increase with growing quantum number $S$.
In our ED calculations, for both the 32- and 36-site clusters, the ground states are located in the singlet sector due to the Lieb-Schultz-Mattis theorem~\cite{Lieb1961}.
With increasing $J_3$, the triplet ($S$ = 1) and quintuplet ($S$ = 2) gaps gradually increase.
At $J_3 \simeq$ 0.275, the lowest-excited singlet state ($S$ = 0) and the lowest quintuplet state cross, and the crossing point is close to the transition point obtained by the vanishing N\'eel order determined by the scaling of the order parameter in Fig.~\ref{fig:MagneticOrders}(a).
At $J_3 \simeq$ 0.4, the lowest-excited singlet state further crosses with the  lowest triplet state.
Although there is a finite-size effect, the NAF-QSL and QSL-PVBS transition points obtained from the level crossings in the low-energy spectra agree with those from the extrapolations of the order parameters.

\begin{figure}[b]
  \centering
  \includegraphics[width=0.49\textwidth]{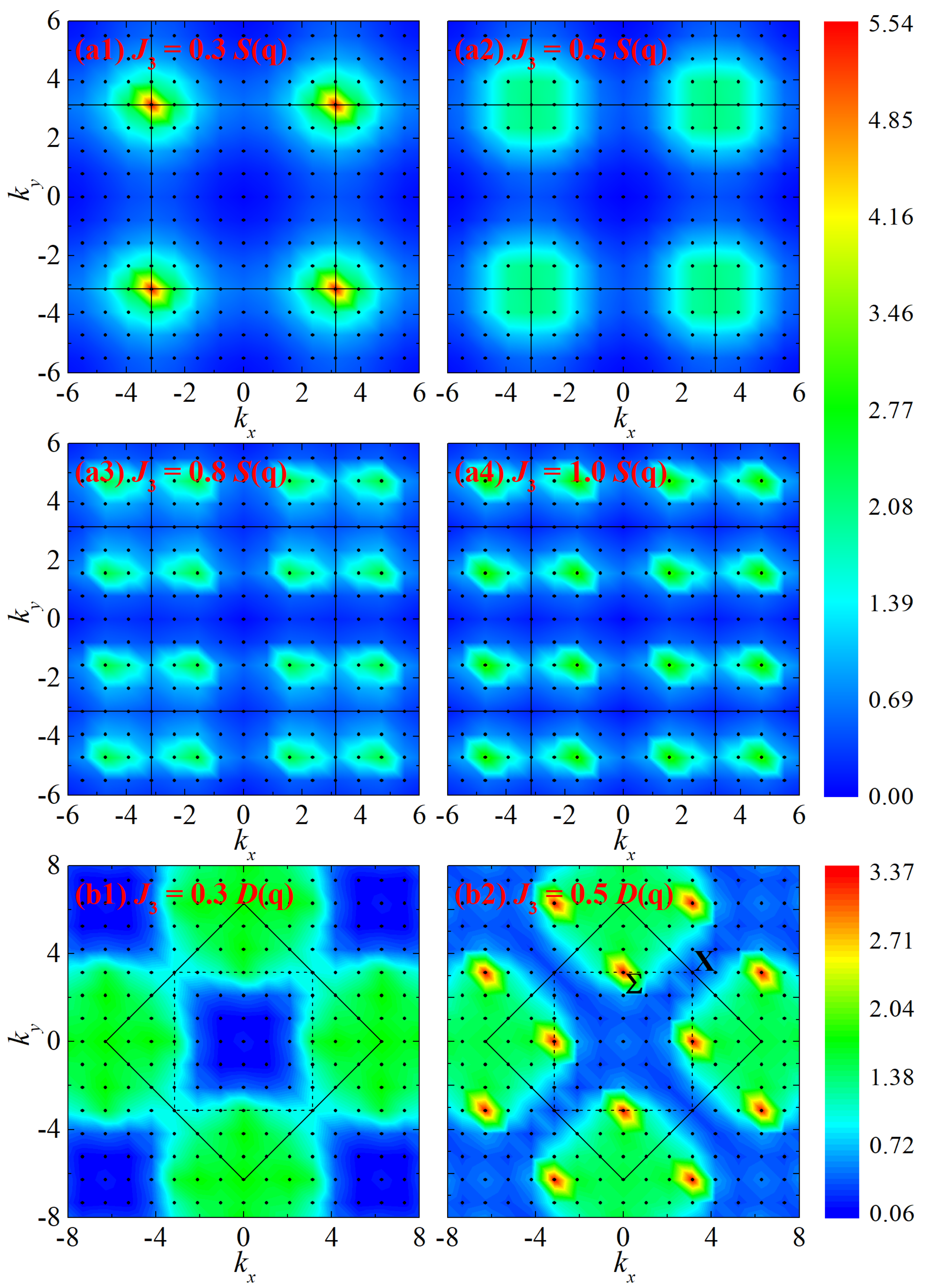}
  \caption{The contour plots of (a1)--(a4) spin structure factor $S(\mathbf{q})$ obtained on the cylinder geometry with $L_y$ = 8 by DMRG and (b1), (b2) dimer structure factor $D(\mathbf{q})$ obtained on the 36-site cluster by ED. The solid lines are the Brillouin zone edge of the original square lattice and the dashed lines in (b1) and (b2) are the Brillouin zone edge of the new square lattice formed by the centers of all the nearest-neighbor bonds in the original lattice.}
  \label{fig:SpinDimerCorr}
\end{figure}

In the PVBS phase, the ground state is a spin-singlet state and has a fourfold degeneracy which can be lifted by spontaneously translational symmetry breaking in the thermodynamic limit, and the triplet excitation would be gapped.
However, due to the finite-size effect in the ED calculation, we can only identify three singlet states of the fourfold degeneracy, as shown in Figs.~\ref{fig:Spectrum}(a) and \ref{fig:Spectrum}(b), which are located in the translational momentum sectors with $k$ = (0, 0), ($\pi$, 0), and (0, $\pi$), respectively.
To determine the low-energy gaps in the PVBS phase, we calculate the energy gaps at $J_3$ = 0.5 obtained on the torus geometry with different lattice sizes, as shown in Fig.~\ref{fig:Spectrum}(c).
The energy gaps of the 32- and 36-site clusters are calculated by ED (see the details in Fig.~\ref{fig:App_Energy_0p5}).
The results of the 8$\times$8 lattice are obtained by DMRG and have been extrapolated with the number of SU(2) states to get the convergent results.
As shown in Fig.~\ref{fig:Spectrum}(c), the singlet gap [$E_1$($S$ = 0) - $E_0$($S$ = 0)] is extrapolated to zero and the triplet gap [$E_0$($S$ = 1) - $E_0$($S$ = 0)] is extrapolated to a finite value, which are consistent with the PVBS state. With the further increase of $J_3$, another (avoided) energy level crossing occurs at $J_3\simeq$  0.725 for the 32--site cluster and at $J_3\simeq$ 0.65 for the 36--site cluster, which indicates the phase transition from the PVBS phase to the spiral order phase.

In Fig.~\ref{fig:Spectrum}(d), we also show the entanglement spectrum obtained by DMRG on the 24$\times$8 cylinder geometry.
From $J_3$ = 0.0 to $J_3$ = 1.0, the lowest spectrum level keeps only onefold degeneracy in the $S$ = 0 sector, which has a gap from the higher spectra.
For the higher spectra above the lowest level, there are crossings between the spectra with $S$ = 0 and $S$ = 1 at $J_3 \simeq $ 0.3 and 0.45, which are close to the phase transition points and may also be taken as the signal of phase transitions.
Moreover, the level crossing also occurs at $J_3 \simeq $ 0.7, which combined with the results of the energy spectra strongly suggest that the phase boundary between the PVBS and the spiral phase locates at $J_3 \simeq $ 0.7.
However, due to the change of the ordering momentum in the spiral phase with growing $J_3$, it is difficult to obtain this transition point from the extrapolations of the spiral order parameter.

\begin{figure*}[t]
  \centering
  \includegraphics[width=\textwidth]{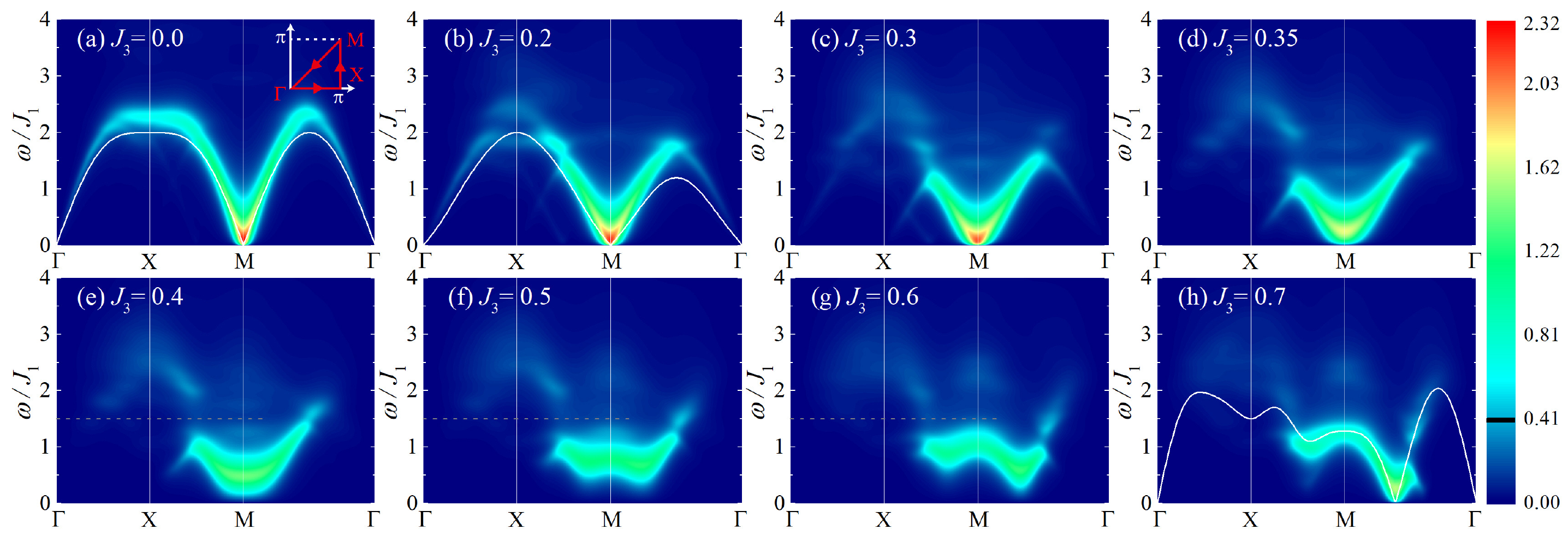}
  \caption{Dynamical spin structure factors $S^{+-}(\mathbf{q},\omega)$ at different $J_3$ calculated by the CPT method with 6 $\times$ 4 cluster tiling. The inset of (a) shows the momentum path in the Brillouin zone. In order to show some high-energy spectra with weaker intensity more clearly, we show $S^{+-}(\mathbf{q},\omega)$ with different mapping to the color bar for the low- and high-intensity excitation spectra. Above the boundary value $U_0$ = 0.4, which is labeled by a black line on the color bar, a logarithmic mapping is used, $U$ = $U_0$ + log$_{10}$[$S^{+-}(\mathbf{q},\omega)$] - log$_{10}$[$U_0$], and $U$ = $S^{+-}(\mathbf{q},\omega)$ while below the boundary value. The Lorentz broadening factor we use for this figure is $\eta$ = 0.15. The white solid lines in (a), (b), and (h) are the dispersion relations obtained by the linear spin-wave theory~\cite{Locher1990, Moreo1990, Rastelli1992, Ferrer1993}. The dashed lines in (e)--(g) demonstrate the separation between the lower-energy and higher-energy excitations in the PVBS phase.}
  \label{fig:CPT-6p4}
\end{figure*}

Employing ED and DMRG, we study the order parameters, energy spectra, and entanglement spectra of the $J_1$--$J_3$ model. The corresponding transition points obtained by different physical observables and lattice sizes are listed in Table~\ref{table:App_TransitionPoint}.
And the final phase diagram is shown in Fig.~\ref{fig:PhaseDiagram}. Besides the NAF phase and spiral phase, which also exist in the classical limit, there are a QSL phase (0.3 $\lesssim J_3 \lesssim 0.375$) and a PVBS phase (0.375 $\lesssim J_3 \lesssim $0.7) in the phase diagram. As shown in Figs.~\ref{fig:MagneticOrders}(a) and \ref{fig:MagneticOrders}(b), the smooth vanishing (appearing) of the extrapolated $m_{s}^{2}$ ($m_{\textrm{VBC}}^{2}$) indicates that the NAF-QSL and QSL-VBS transitions may both be continuous phase transitions. The PEPS simulation results even suggested that these two transitions may belong to new types of universality classes~\cite{WYLiu2021tri3}. Nonetheless, weak first-order transitions cannot be excluded due to the limit of the studied system size.

To further characterize the different phases, we also show the static spin structure factor $S(\mathbf{q})$ and the dimer structure factor $D(\mathbf{q})$ at different couplings in Fig.~\ref{fig:SpinDimerCorr}.
At both $J_3$ = 0.3 and 0.5, the static spin structure factors $S(\mathbf{q})$ show broad peaks around the $M$ point, which come from the short-range spin correlation inherited from the N\'eel order and become weaker with the increase of $J_3$. When approaching the spiral phase, the wave vector of maximum $S(\mathbf{q})$ gradually deviates from the $M$ point. By using field-theory techniques and series expansion, a previous study~\cite{Kharkov2018} found that the spiral order is established at $J_3 \simeq$ 0.55. In our study, the long-range spiral order is suppressed by quantum fluctuation until $J_3 \simeq$ 0.7. However, there would be some short-range spiral order existing for 0.55 $\lesssim J_3 \lesssim $0.7 and how the short-range spiral order will affect the PVBS state still needs further study.
In the spiral phase, the wave vector ($Q$, $Q$) of the Bragg peak should keep approaching ($\pi$/2, $\pi$/2) with increasing $J_3$.
As shown in Figs.~\ref{fig:SpinDimerCorr}(a3) and \ref{fig:SpinDimerCorr}(a4), although the 24$\times$8 cylinder may not have ($Q$, $Q$) in the momentum space, the broadened peaks near ($\pi$/2, $\pi$/2) are consistent with the spiral order on such a finite cluster.
For the dimer structure factor $D(\mathbf{q})$ of the 36-site cluster obtained by ED, there is a wide range of broad continua around the original and new Brillouin zone edge at $J_3$ = 0.3.
On the other hand, $D(\mathbf{q})$ has sharp peaks at the $\Sigma$ momentum point of the new Brillouin zone at $J_3$ = 0.5, which agrees with the emergent VBS order.

\subsection{Magnetic Excitation}
\label{Sec:Excitation}

\begin{figure}[t]
  \centering
  \includegraphics[width=0.49\textwidth]{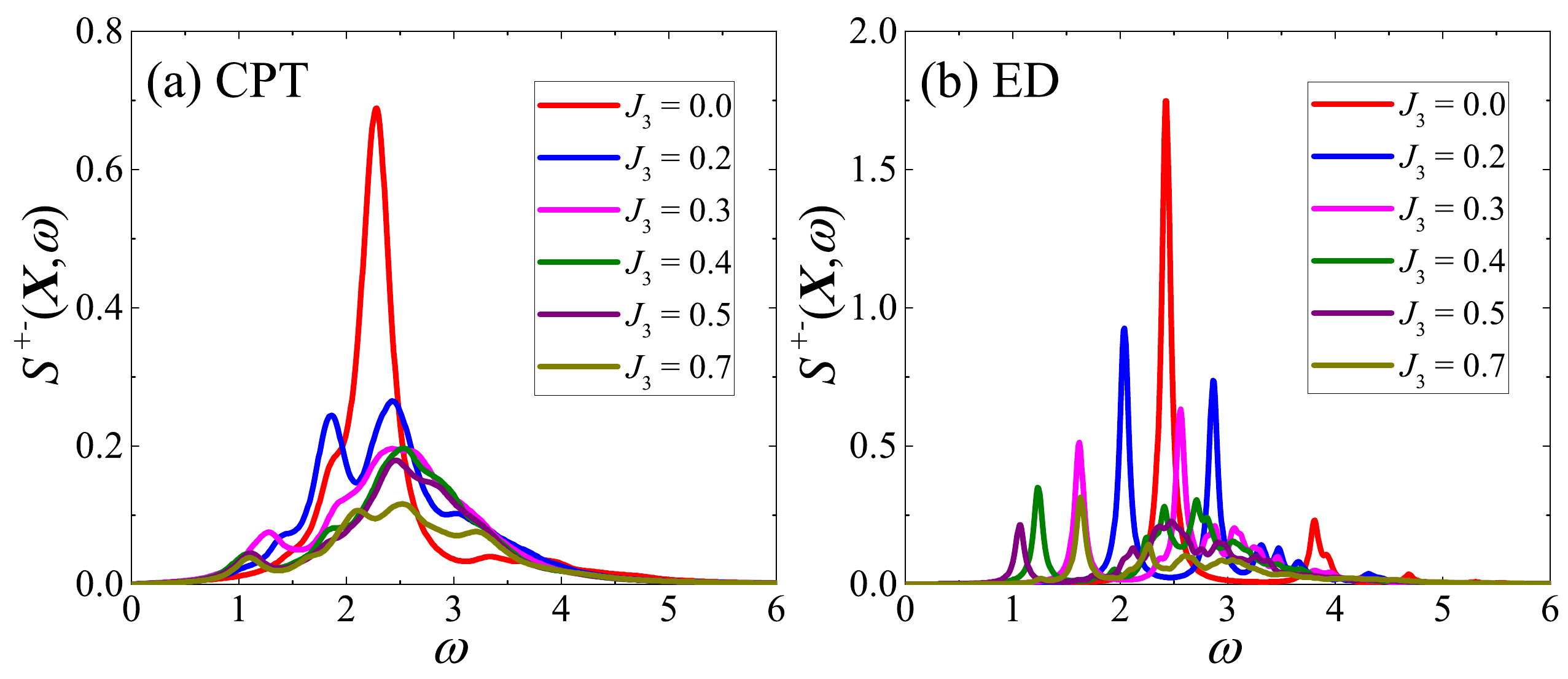}
  \caption{Dynamical spin structure factors $S^{+-}(\mathbf{q},\omega)$ at the $X$ momentum point with different $J_3$, which are calculated by (a) CPT and (b) ED.}
  \label{fig:TD_J3_X}
\end{figure}

In this section, we study the dynamical spin structure factor by using the CPT and ED methods.
In Fig.~\ref{fig:CPT-6p4}, we show $S^{+-}(\mathbf{q},\omega)$ obtained by the CPT method along the high-symmetry path $\Gamma (0, 0) \rightarrow X (\pi, 0) \rightarrow M (\pi, \pi) \rightarrow \Gamma (0, 0)$ of the Brillouin zone [see the inset of Fig.~\ref{fig:CPT-6p4}(a)].
In the N\'eel phase, the spontaneous breaking of spin SU(2) symmetry contributes a gapless Goldstone mode at the $M$ point, as shown in Figs.~\ref{fig:CPT-6p4}(a) and \ref{fig:CPT-6p4}(b). At low energy, our CPT results are consistent with the dispersion relations obtained by the linear spin-wave theory~\cite{Locher1990, Moreo1990, Rastelli1992, Ferrer1993}, which indicates the reliability of our CPT results. At high energy around $\omega \approx$ 2.0, the excitations obtained by the two methods have a slight difference which may be induced by magnon interactions and quantum fluctuation. And considering the first-order or even second-order 1/$S$ correction in the spin-wave theory would reduce the difference~\cite{Majumdar2010}.
For $J_3$ = 0.0 in Fig.~\ref{fig:CPT-6p4}(a), our CPT data are consistent with the previous CPT results~\cite{ShLYu2018}.
With growing $J_3$, the high-energy excitation at the $X$ point becomes weaker and splits into two branches.
Meanwhile, the intensity of the excitation tail slightly enhances.
As shown in Fig.~\ref{fig:TD_J3_X}(b), these features can also be observed in the ED results obtained on the 36-site cluster, although the ED results have sharper peaks due to stronger finite-size effect and small Lorentz broadening factor $\eta$.
With growing $J_3$, for example at $J_3$ = 0.2, there is another branch of the excitation around the $M$ point emerging at the energy scale $\omega \approx$ 2.0.
Meanwhile, the excitation mode at around point ($\pi$/2, $\pi$/2) shifts to lower energy. As shown in Fig.~\ref{fig:CPT-6p4}(b), the deviation at high energy around $\omega \approx$ 2.0 is more obvious with increasing $J_3$, like the split of the excitation at the $X$ point and the emergence of another branch of excitation around the $M$ point at $\omega \approx$ 2.0, and this may also indicate the enhancement of the quantum fluctuation when approaching the QSL phase. For the N\'eel phase in the $J_1$--$J_2$ Heisenberg model, the gapless Goldstone mode remains at around the $M$ point~\cite{ShLYu2018}. With growing $J_2$, the excitation mode at around the $X$ point gradually reduces to lower energy without visible splitting and, finally, the whole spectrum turns into a broad continuum in the QSL phase~\cite{ShLYu2018}.

\begin{figure}[b]
  \centering
  \includegraphics[width=0.49\textwidth]{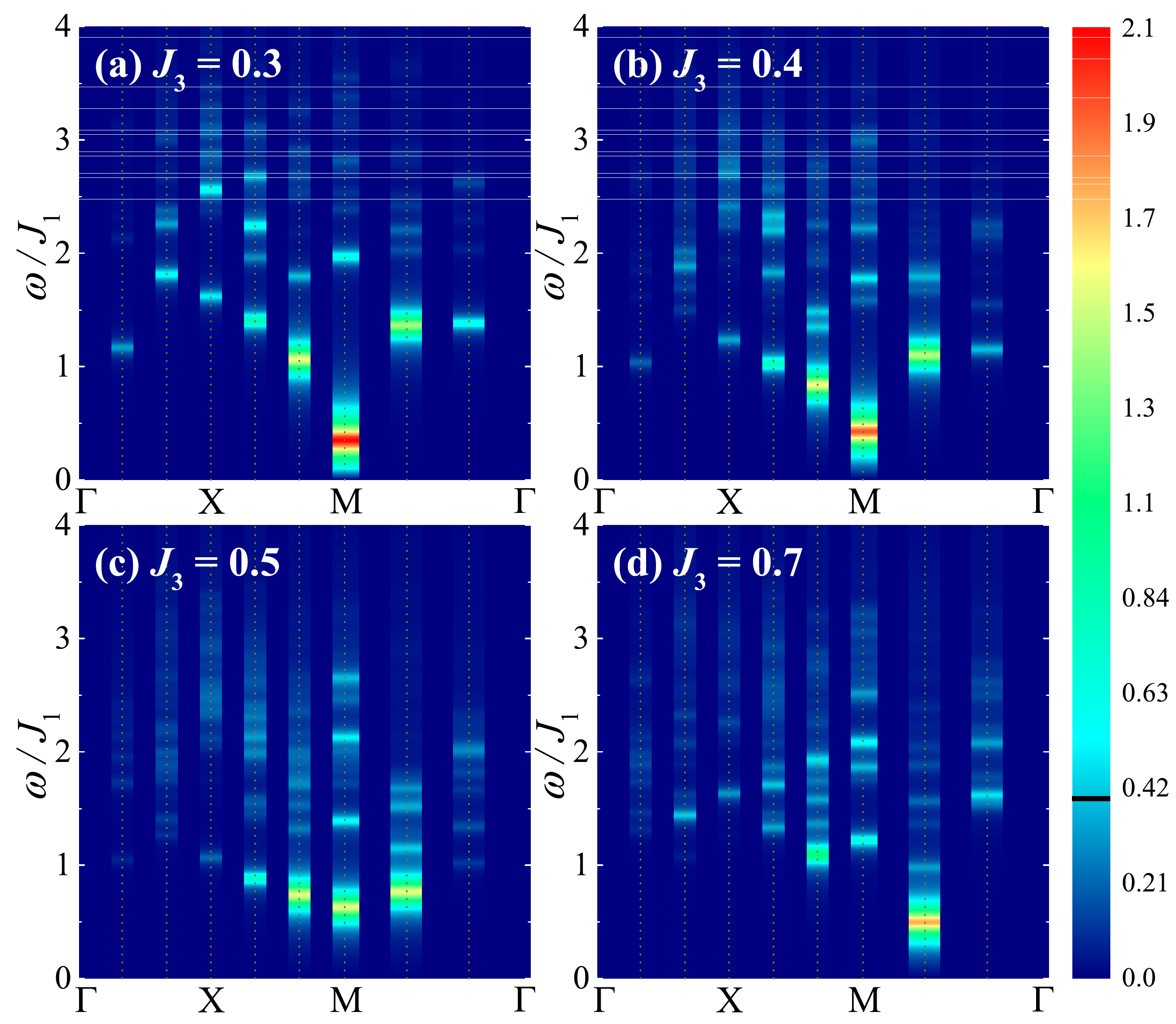}
  \caption{Dynamical spin structure factors $S^{+-}(\mathbf{q},\omega)$ at different $J_3$ calculated by ED. The results are shown in the similar way to Fig.~\ref{fig:CPT-6p4} and the boundary value $U_0$ = 0.4 is labeled by a black line on the color bar. The Lorentz broadening factor we use for this figure is $\eta$ = 0.05. The dotted lines indicate the momentum points contained in the momentum space of the 36-site cluster.}
  \label{fig:SpmTD-6p6}
\end{figure}

In the QSL phase of the $J_1$--$J_3$ model, the strongest intensity still locates at around the $M$ point, as shown in Figs.~\ref{fig:CPT-6p4}(c) and \ref{fig:CPT-6p4}(d). At $J_3$ = 0.3, the lower branch of excitations at the $X$ point becomes weaker and the corresponding excitation gap is smaller, compared with the spectra in the N\'eel phase. For the higher branch at the $X$ point, it turns into some broad continua with only slightly weakened spectral intensity. As shown in  Fig.~\ref{fig:SpmTD-6p6}(a) and Fig.~\ref{fig:TD_J3_X}(b), these dispersive excitations can also be seen in the ED results for $\omega \gtrsim$ 2.25 and are formed by many weak excitation peaks. These continua may attribute to the deconfined spinons in the QSL phase. At $J_3$ = 0.35, a weak signal of a small gap is found in the low-energy spectrum and one can also see some weak continua in high energy around the $M$ point. However, by using ED as a solver for the CPT method, we calculate the clusters up to 24 sites, in which it may still be hard to identify the gapped or gapless nature of the QSL phase. For the QSL phase in the $J_1$--$J_2$ Heisenberg model, the spectrum is characterized by a broad continuum~\cite{ShLYu2018}, while in the $J_1$--$J_3$ model, there are only some weak continua in the low-energy spectrum around the $X$ point when $J_3$ = 0.3 and 0.35.

Figures~\ref{fig:CPT-6p4}(e)--\ref{fig:CPT-6p4}(g) show the dynamical spin structure factors $S^{+-}(\mathbf{q},\omega)$ obtained by CPT in the PVBS phase, which have the gapped spin-triplet excitation. Interestingly, it seems that there is another excitation with residual broad continua above $\omega \sim 1.5 J_1$, which is separated from the lower-energy excitation by a gap [see the dashed lines in Figs.~\ref{fig:CPT-6p4}(e)--\ref{fig:CPT-6p4}(g)].
Such a similar feature has also been observed in the quantum Monte Carlo results of the $2 \times 2$ checkerboard model~\cite{YXu2019}. And we also show our CPT results of the same model in Fig.~\ref{fig:App_Checkerboard} to do the comparison. In that model, caused by the nonuniform interactions, the translational symmetry is already broken in the Hamiltonian and its ground state is in the plaquette phase when the ratio of the inter- to the intra-plaquette interaction is smaller than 0.548524(3)~\cite{XRan2019}. The plaquette state can adiabatically connect to a state which is formed by the direct product of the four-site plaquettes. In the plaquette phase, the two excitations locating at the low and high energy are owing to two different kinds of triplet excitations originating from the four-site plaquette, which are shown in Fig.~\ref{fig:App_Plaquette}. In the $2 \times 2$ checkerboard model, without interplaquette interaction, these two kinds of excitations are localized. After adding a weak interplaquette interaction which can be seen as a perturbation, the quasi-particle of the local excitations can move in the whole lattice and the lower-energy branch turns into a dispersive excitation. Compared with the lower-energy branch, the high-energy branch is more localized due to the much larger excitation gap and the effective interactions are very weak, which cause the high-energy branch to be nearly flat. In the $J_1$--$J_3$ model, we can also ascribe the two excitations observed in Figs.~\ref{fig:CPT-6p4}(e)--\ref{fig:CPT-6p4}(g) to these two kinds of triplet excitations. Nevertheless, there are still some differences between the spectra of the two models. As shown in Fig.~\ref{fig:CPT-6p4}(e), the low-energy excitation at $J_3$ = 0.4 is a gapped triplon excitation around the $M$ point, which is very similar to the one in the $2 \times 2$ checkerboard model. But for $\omega \gtrsim 1.5 J_1$, the high-energy branch is a dispersive excitation with weaker intensity. In $J_1$--$J_3$ model, the PVBS phase is formed by spontaneously breaking the translational symmetry and the local excitations at high energy have higher mobility due to the uniform $J_1$ and $J_3$ interactions. And these lead to the differences between the spectra of two models. As shown in Fig.~\ref{fig:SpmTD-6p6}(c), two triplet excitations can also be seen in the ED results, although the high-energy dispersion is not so clear due to the finite-size effect. Because the PVBS ground state is a singlet state with the translational wave vector at $\Gamma$ point and $S^{+-}(\mathbf{q},\omega)$ mainly captures the excitation with $\Delta S$ = 1, the low-energy excitations shown in Fig.~\ref{fig:SpmTD-6p6}(c) correspond to the triplet excitations, especially around the $M$ point in Fig.~\ref{fig:App_Energy_0p5}(b). With growing $J_3$ in the PVBS phase, the excitation gap at the $M$ point is enlarged and the minimum gap of the whole Brillouin zone gradually transforms from the $M$ point to another wave vector along the $M \rightarrow \Gamma$ path. And these changes may be caused by the short-range spiral order. As shown in Fig.~\ref{fig:CPT-6p4}(h), when the spiral order is established at $J_3$ = 0.7, a new gapless magnon mode develops around ($Q$, $Q$) which is located between the $M$ and $\Gamma$ points. The good agreement between the CPT results and the spin-wave dispersion relation also indicates the existence of the spiral order at $J_3$ = 0.7. And the small excitation gap of the ED result in Fig.~\ref{fig:SpmTD-6p6}(d) is caused by the finite-size effect.

\section{Summary and Discussion}
\label{Sec:Summary}

By using the DMRG and ED methods, we calculate the order parameters and the low-lying energy and entanglement spectra of the spin-$1/2$ $J_1$--$J_3$ Heisenberg model on the square lattice.
We obtain the phase diagram of this model and find a paramagnetic region ($0.3 \lesssim J_3/J_1 \lesssim 0.7$) sandwiched between the N\'eel and the spiral order phases.
For $0.375 \lesssim J_3/J_1 \lesssim 0.7$, a PVBS phase is characterized by our numerical calculation, including the extrapolated results of the dimer order parameters $m_{\textrm{VBC}} \neq 0$ and $m_{\textrm{Col}} = 0$ as well as the vanished (finite) spin-singlet gap (spin-triplet gap) in the thermodynamic limit. For $0.3 \lesssim J_3/J_1 \lesssim 0.375$, we find absent N\'eel and dimer order, which agree with a QSL phase.

Furthermore, we use the CPT and ED methods to study the dynamical spin structure factor $S^{+-}(\mathbf{q},\omega)$.
While a Goldstone mode appears at the $M$ point in the N\'eel phase, a new gapless magnon mode developing around ($Q$, $Q$) is found in the spiral order phase, which approaches ($\pi/2, \pi/2$) with growing $J_3$.
In the QSL phase, some dispersive weak excitations around the $X$ and $M$ points are captured by our results.
By comparing with the dynamical spin structure factor of the 2$\times$2 checkerboard model, the CPT method is able to capture most of the characteristics of the PVBS phase.
Similar to the plaquette phase of the 2$\times$2 checkerboard model, except for the triplet excitation around the $M$ point at low energy, there is another triplet excitation surrounding with some continua at high energy. The difference between the excitation spectra of these two models may be owing to the spontaneously translational symmetry breaking of the PVBS phase and the uniform interactions in the $J_1$--$J_3$ Heisenberg model. In addition, except for the $J$--$Q_6$ model~\cite{Takahashi2020} and the $J_1$--$J_2$--$J_3$ Heisenberg model~\cite{Capriotti2000, Mambrini2006, SSGong2014, Murg2009, JFYu2012, Reuther2011, Daniel2021}, the four-fold degenerate PVBS phase for the square lattice has been rarely found in models without the explicit translational symmetry breaking in the Hamiltonian. Our results will also provide more theoretical understanding of the magnetic excitation of such a PVBS phase.

\begin{acknowledgments}

M.W. thanks Shun-Li Yu for helpful discussions on the CPT methods and Caiyuan Ye for helping to improve the ED codes. We thank Wen-Yuan Liu for helpful discussions. This work is supported by NKRDPC-2017YFA0206203, NKRDPC-2018YFA0306001, NSFC-11804401, NSFC-11974432, NSFC-11832019, NSFC-11874078, NSFC-11834014, NSFC-92165204, GBABRF-2019A1515011337, Leading Talent Program of Guangdong Special Projects (201626003), Shenzhen International Quantum Academy (SIQA202102), and Guangzhou Basic and Applied Basic Research Foundation.

\end{acknowledgments}

\appendix

\begin{figure}[t]
  \centering
  \includegraphics[width=0.45\textwidth]{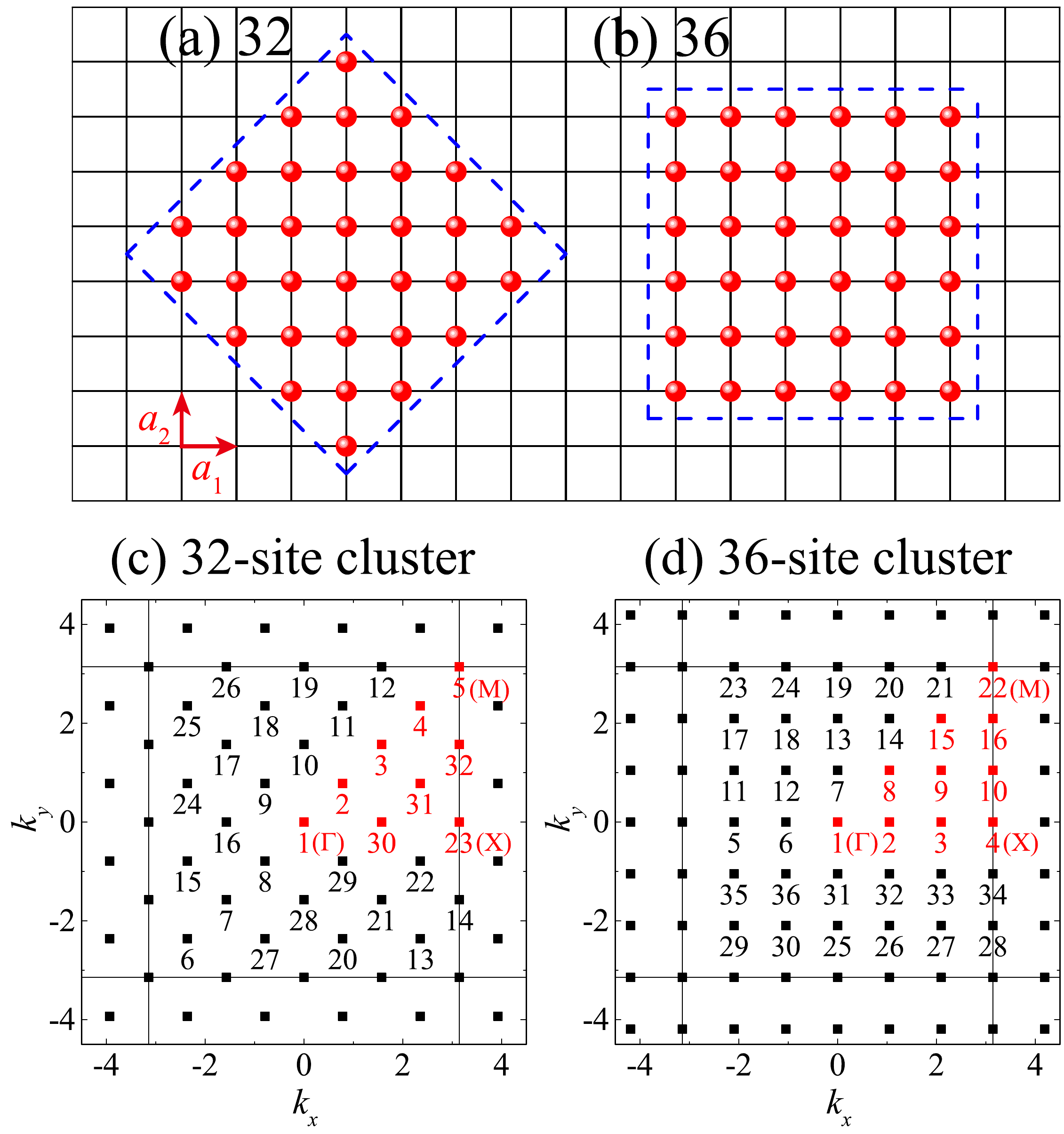}
  \caption{(a),(b) The finite-size clusters used in the ED calculations. $\mathbf{a}_{1}=(a,0)$ and $\mathbf{a}_{2}=(0,a)$ are primitive vectors of the square lattice. (c),(d) The momentum points of the 32- and 36-site clusters.}
  \label{fig:App_Lattice}
\end{figure}

\section{Finite-size Clusters}
\label{App:FSClusters}

In this paper, we use the Lanczos ED to calculate the energy spectrum, static dimer structure factor, and dynamical spin structure factor.
By using the translational symmetry, spin-inversion symmetry, and sparse matrix storage technique to reduce the cost of the memory and speed up the calculation, we can do the diagonalizations up to 36 sites under the periodic boundary conditions (see Fig.~\ref{fig:App_Lattice}).
Due to the $C_{4v}$ symmetry of the lattice, we can only calculate the energy spectrum of the momentum points, labeled as red colors in Figs.~\ref{fig:App_Lattice}(c) and \ref{fig:App_Lattice}(d), instead of all 32 or 36 momentum points.
The eigenstates at the other momentum points have the degenerate energy with the eigenstates at these red momentum points. Both the $M$ = ($\pi$, $\pi$) and $X$ = ($\pi$, 0), (0, $\pi$) momentum points are contained in these two clusters.
The $M$ momentum point is important to capture both the N\'eel order and the VBS order.
Along the high-symmetry path $\Gamma \rightarrow X \rightarrow M \rightarrow \Gamma$ on the 36-site cluster, there are nine different momentum points which are helpful to get more details of the excitation spectrum on finite-size clusters and to compare with the results obtained by the CPT.

\section{More Details On Identifying Phase Transition Points}
\label{App:TransitionPoint}

In Sec.~\ref{Sec:PhaseDiagram}, as shown in Figs.~\ref{fig:MagneticOrders} and \ref{fig:Spectrum}, we obtain the phase transition points of the $J_1$--$J_3$ model by extrapolating the order parameters and studying the level crossings in the energy and entanglement spectra. In Table~\ref{table:App_TransitionPoint}, we list all the transition points obtained by different physical observables as well as the corresponding lattice size and method used in the calculation.

As shown in Fig.~\ref{fig:PhaseDiagram}, there are three transition points in the phase diagram of the $J_1$--$J_3$ model, including the NAF-QSL, QSL-PVBS, and PVBS-Spiral transition. And after considering all the results listed in Table~\ref{table:App_TransitionPoint}, we identify that these three phase transitions occur at around $J_3 \simeq$ 0.3, 0.375, and 0.7, respectively. And except for the NAF-QSL transition, which has an error of 0.025, we estimate that the errors of the other two transition points are 0.05.

\begin{figure}[t]
  \centering
  \includegraphics[width=0.49\textwidth]{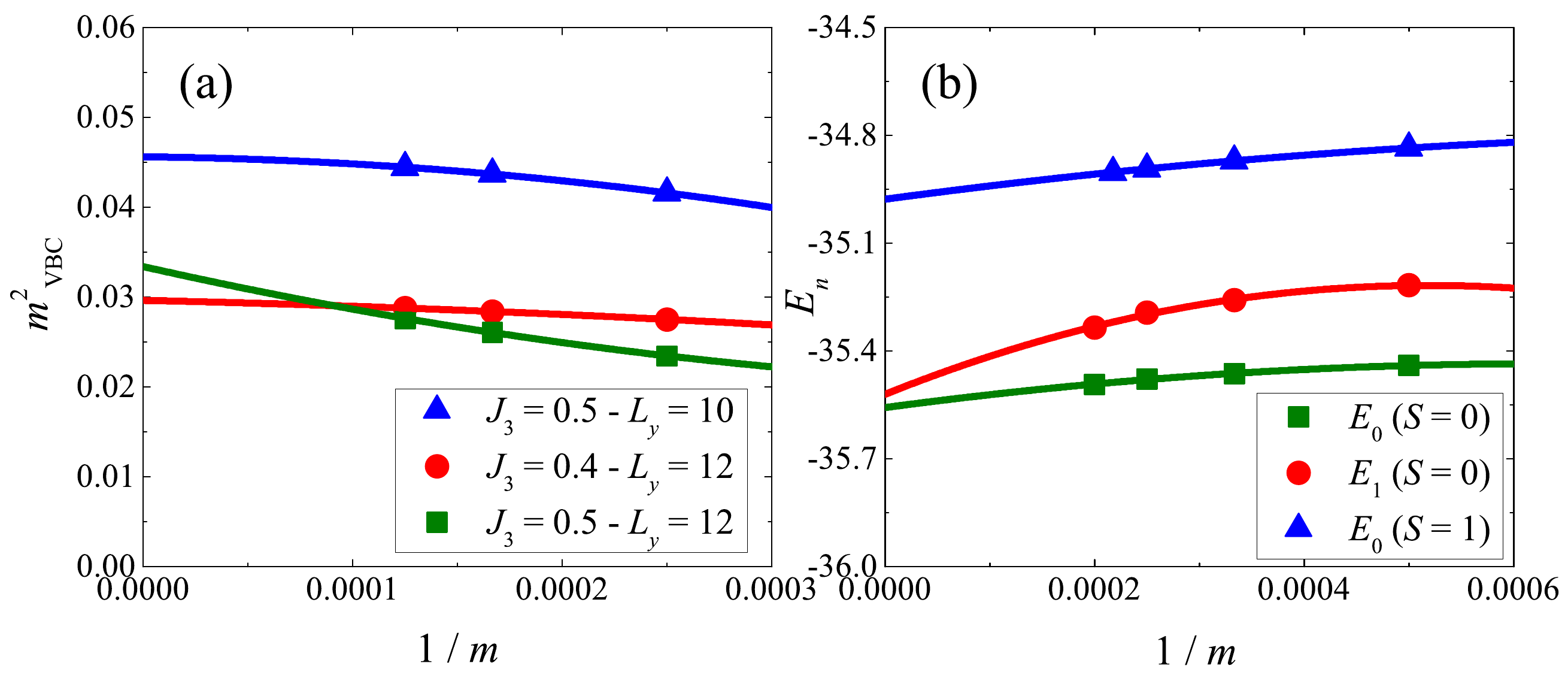}
  \caption{(a) The second-order polynomial extrapolation of $m_{\textrm{VBC}}^{2}$ at different $J_3$ on cylinder geometries with $L_y$ = 10 and 12 vs $1/m$. (b) The second-order polynomial extrapolation of the energy at $J_3$ = 0.5 obtained on the 8$\times$8 cluster under the periodic boundary conditions.}
  \label{fig:App_Fit}
\end{figure}

\begin{figure}[t]
  \centering
  \includegraphics[width=0.49\textwidth]{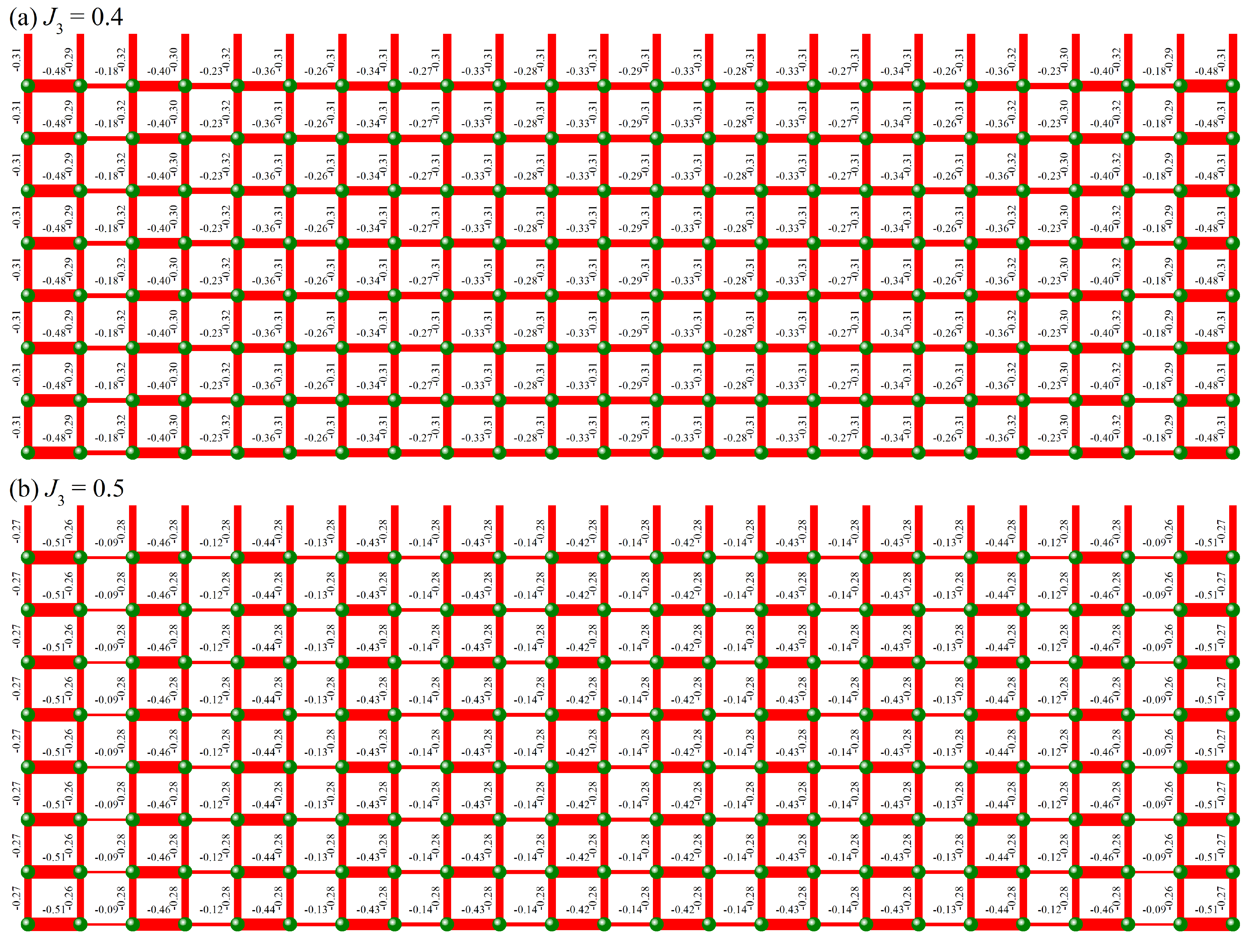}
  \caption{The nearest-neighbor bond energy obtained on the  cylinder geometry with $L_y = 8, L_x = 24$ at $J_3$ = (a) 0.4 and (b) 0.5, which are calculated by the DMRG.}
  \label{fig:APP_BondEnergy_Cylinder}
\end{figure}

\begin{figure}[b]
  \centering
  \includegraphics[width=0.49\textwidth]{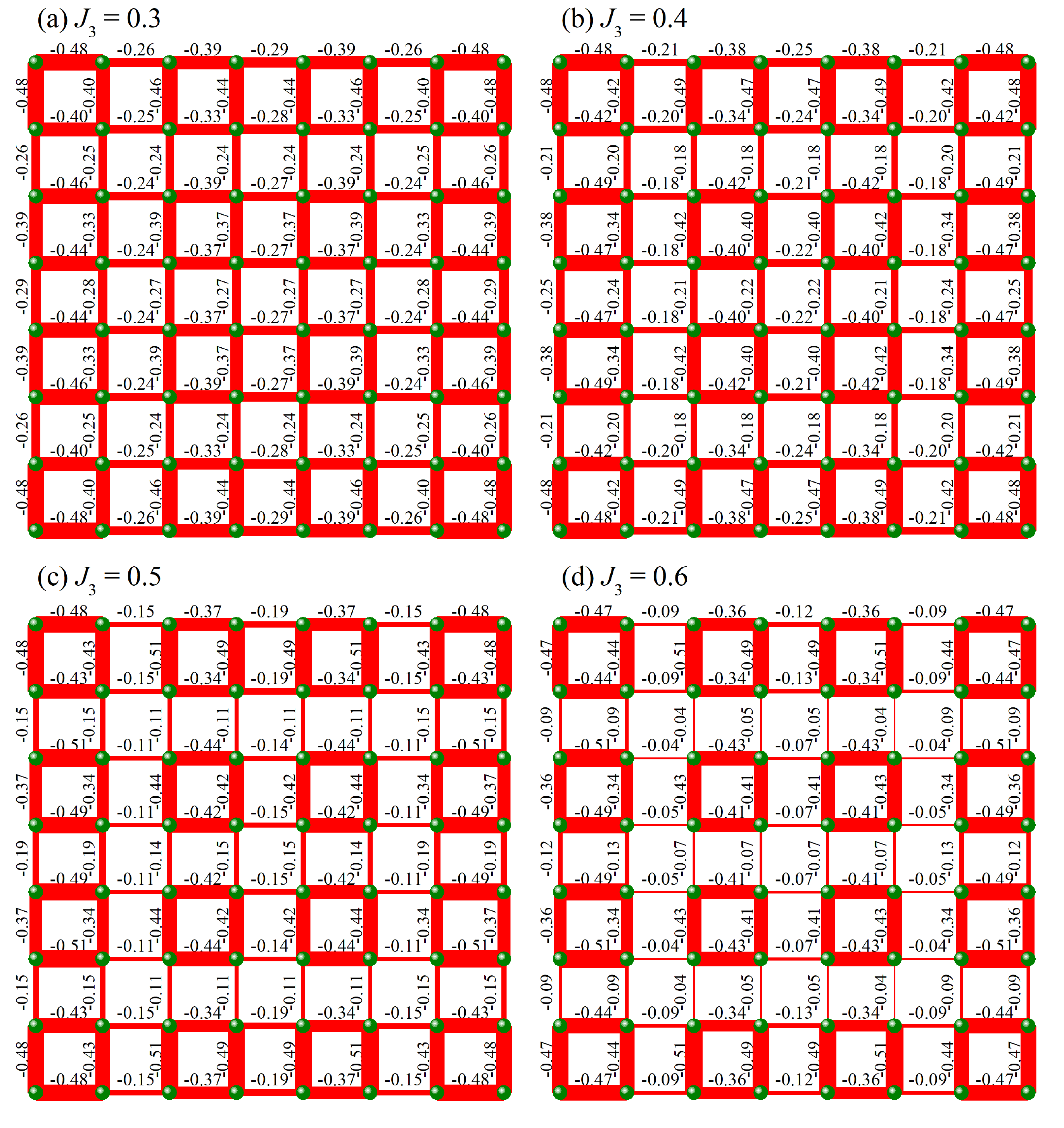}
  \caption{The nearest-neighbor bond energy on the 8$\times$8 cluster with the open boundary conditions at $J_3$ = (a) 0.3, (b) 0.4, (c) 0.5, and (d) 0.6, which are calculated by the DMRG.}
  \label{fig:APP_BondEnergy}
\end{figure}

\section{More details of the DMRG results}
\label{App:DMRG}

In the main text, we show the size extrapolation of the dimer order parameter $m_{\textrm{VBC}}^{2}$ in Fig.~\ref{fig:MagneticOrders}(b).
In order to obtain accurate results, we have extrapolated $m_{\textrm{VBC}}^{2}$ versus DMRG bond dimensions $1/m$ before size scaling.
Due to the limit of computational cost, we can only keep up to 8000 SU(2) states in DMRG calculation.
As shown in Fig.~\ref{fig:App_Fit}(a), 8000 SU(2) states are almost enough for the results of $m_{\textrm{VBC}}^{2}$ at $J_3$ = 0.5 with $L_y$ = 10 and $J_3$ = 0.4 with $L_y$ = 12.
For larger $J_3$ or $L_y$, it seems that more states are needed to ensure complete convergence.
This is one reason for the difficulty to determine the phase boundary using the scaling of the dimer order parameter.
Another reason is that the extrapolation of $m_{\textrm{VBC}}^{2}$ obtained on the cylinder geometry may not describe the VBS order well when $J_3$ is large and deep inside the PVBS phase. As shown in Fig.~\ref{fig:APP_BondEnergy_Cylinder}, by calculating on the cylinder geometry which has broken the translational symmetry in the $x$ direction, the boundary-induced dimer order decays fast from the boundary to the bulk, and the bond energy is relatively uniform in the bulk at $J_3$ = 0.4. But at $J_3$ = 0.5, the bond energy strongly breaks the translational symmetry and keeps highly dimerized even in the bulk, which will lead to the failure of identifying the dimer order by $m_{\textrm{VBC}}^{2}$. Nonetheless, when $J_3\leq 0.4$, the extrapolation of $m_{\textrm{VBC}}^{2}$ can still describe the VBS order well and the extrapolation versus $1/m$ can give more reliable results, which are important to identify the PVBS order.
In Fig.~\ref{fig:App_Fit}(b), we also show the bond dimension scaling of the singlet gap and triplet gap of the 8$\times$8 cluster under the periodic boundary conditions at $J_3$ = 0.5, by keeping up to 5000 and 4600 SU(2) states, respectively.

\begin{figure}[b]
  \centering
  \includegraphics[width=0.49\textwidth]{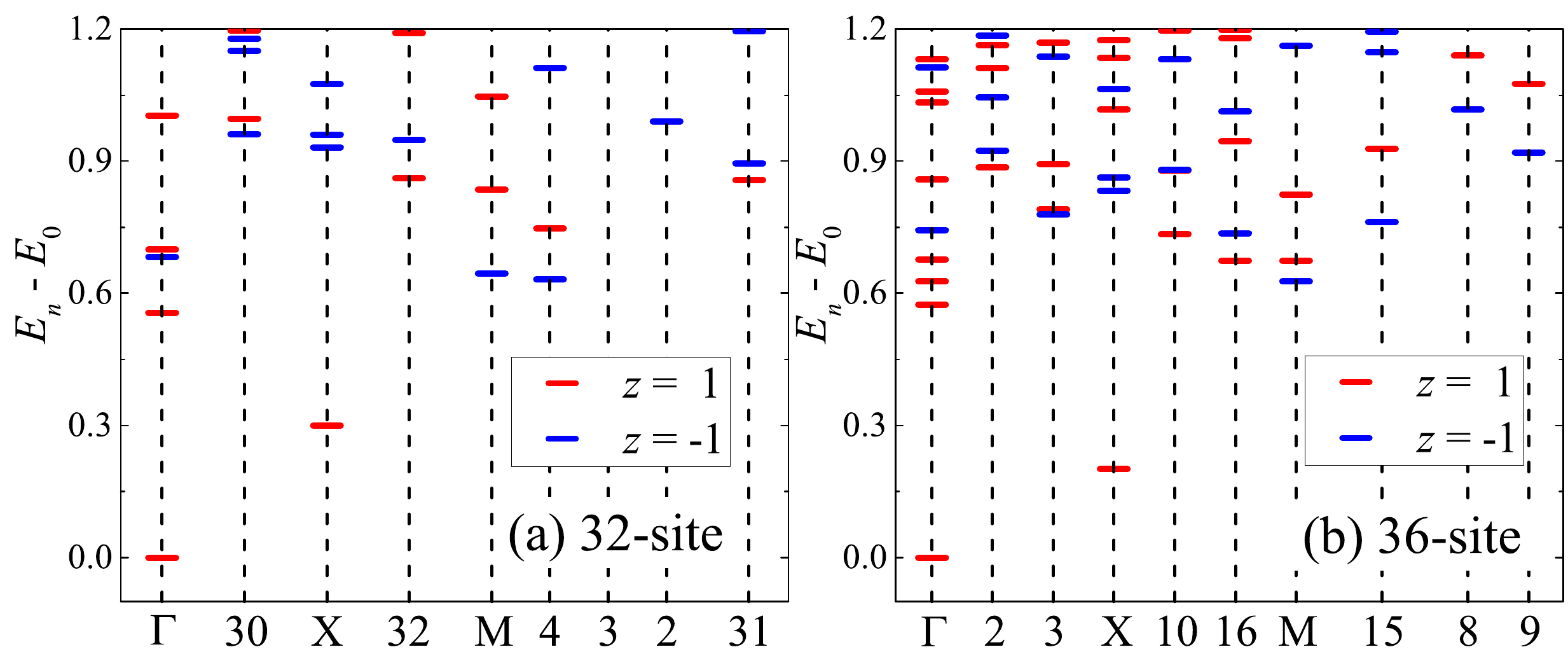}
  \caption{The low-energy spectra with different momentum points at $J_3$ = 0.5 obtained on the (a) 32-site and (b) 36-site clusters. The red and blue bars represent energies with $z$ = 1 and -1, where $z$ is the eigenvalue of the spin-inversion operator $Z$~\cite{Sandvik2010}.}
  \label{fig:App_Energy_0p5}
\end{figure}

In order to have an intuition of the PVBS phase, we show the nearest-neighbor bond energy on the 8$\times$8 cluster with the open boundary conditions in Fig.~\ref{fig:APP_BondEnergy}.
For the cluster with the periodic boundary conditions, because of the translational symmetry the nearest-neighbor bond energies are uniform.
With the broken translational symmetry in the open boundary conditions, the nearest-neighbor bond energy shows an obvious plaquette pattern at $J_3$ = 0.4, 0.5, and 0.6, which becomes sharper with the increase of $J_3$.
Our DMRG result at $J_3$ = 0.5, as shown in Fig.~\ref{fig:APP_BondEnergy}(c), agrees well with the result obtained by the PEPS~\cite{Murg2009}.
On the 8$\times$8 cluster, every four neighbor sites form a plaquette. The nearest-neighbor bond energies inside the plaquette are strong and the interplaquette bond energies are much weaker, which is consistent with a PVBS phase.
This plaquette pattern is much weaker at $J_3$ = 0.3, as shown in Fig.~\ref{fig:APP_BondEnergy}(a), and in the N\'eel phase the nearest-neighbor bond energies only have slight differences with no sign of the plaquette pattern.

\section{Energy Spectrum in the ED calculation}
\label{App:EDSpectrum}

\begin{figure}[htp!]
  \centering
  \includegraphics[width=0.48\textwidth]{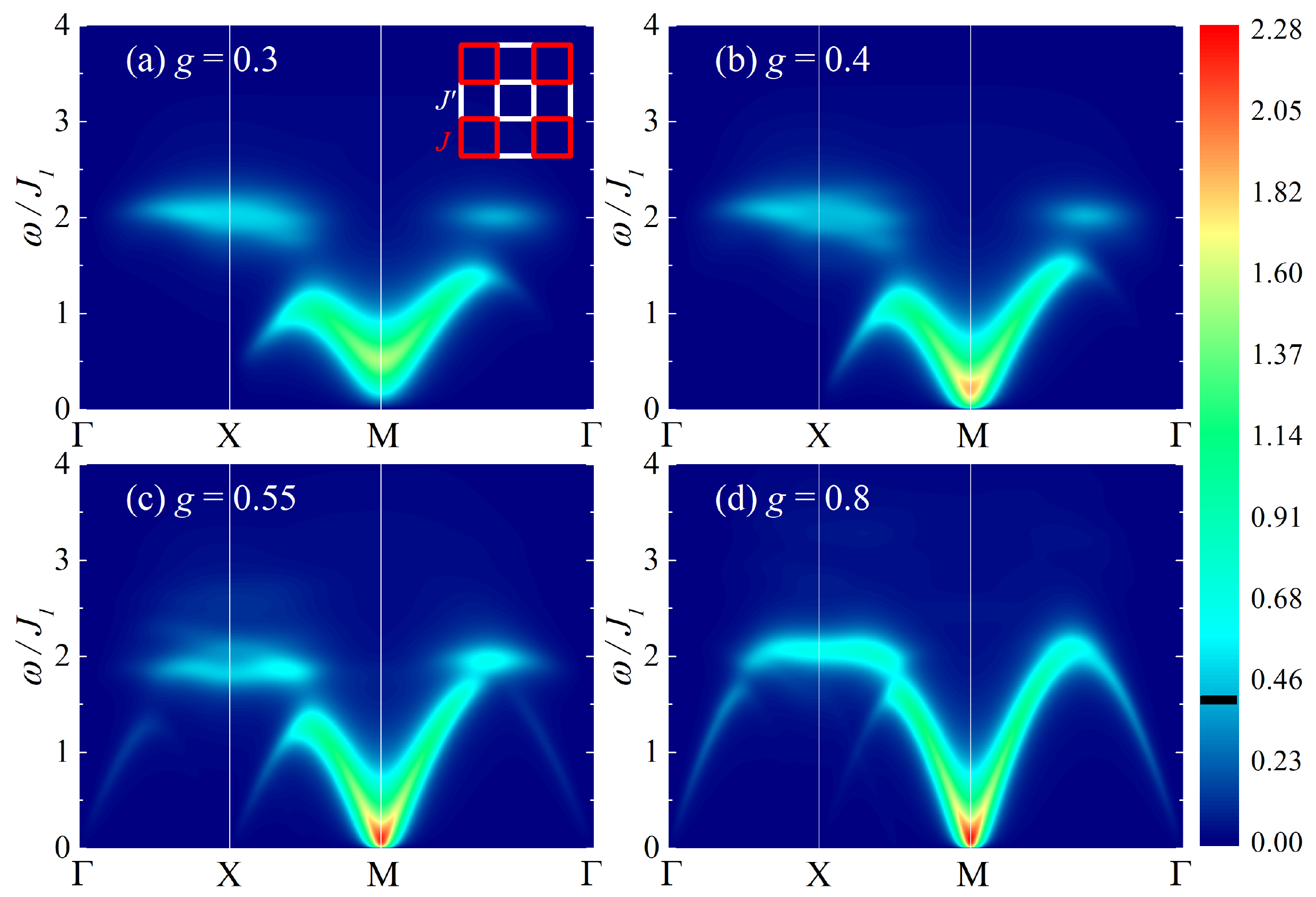}
  \caption{The dynamical spin structure factors $S^{\pm}(\mathbf{q},\omega)$ for the 2$\times$2 checkerboard model with different $g$ calculated by CPT. The inset of (a) shows the structure of the 2$\times$2 checkerboard model in which the red and white lines represent the intraplaquette interaction $J$ and interplaquette interaction $J'$, respectively. And the parameter $g$ is defined as $g = J' / J$. The results are shown in the similar way to Fig.~\ref{fig:CPT-6p4} and the boundary value $U_0$ = 0.4 is labeled by a black line on the color bar. The Lorentz broadening factor we use for this figure is $\eta$ = 0.15.}
  \label{fig:App_Checkerboard}
\end{figure}

\begin{figure}[htp!]
  \centering
  \includegraphics[width=0.47\textwidth]{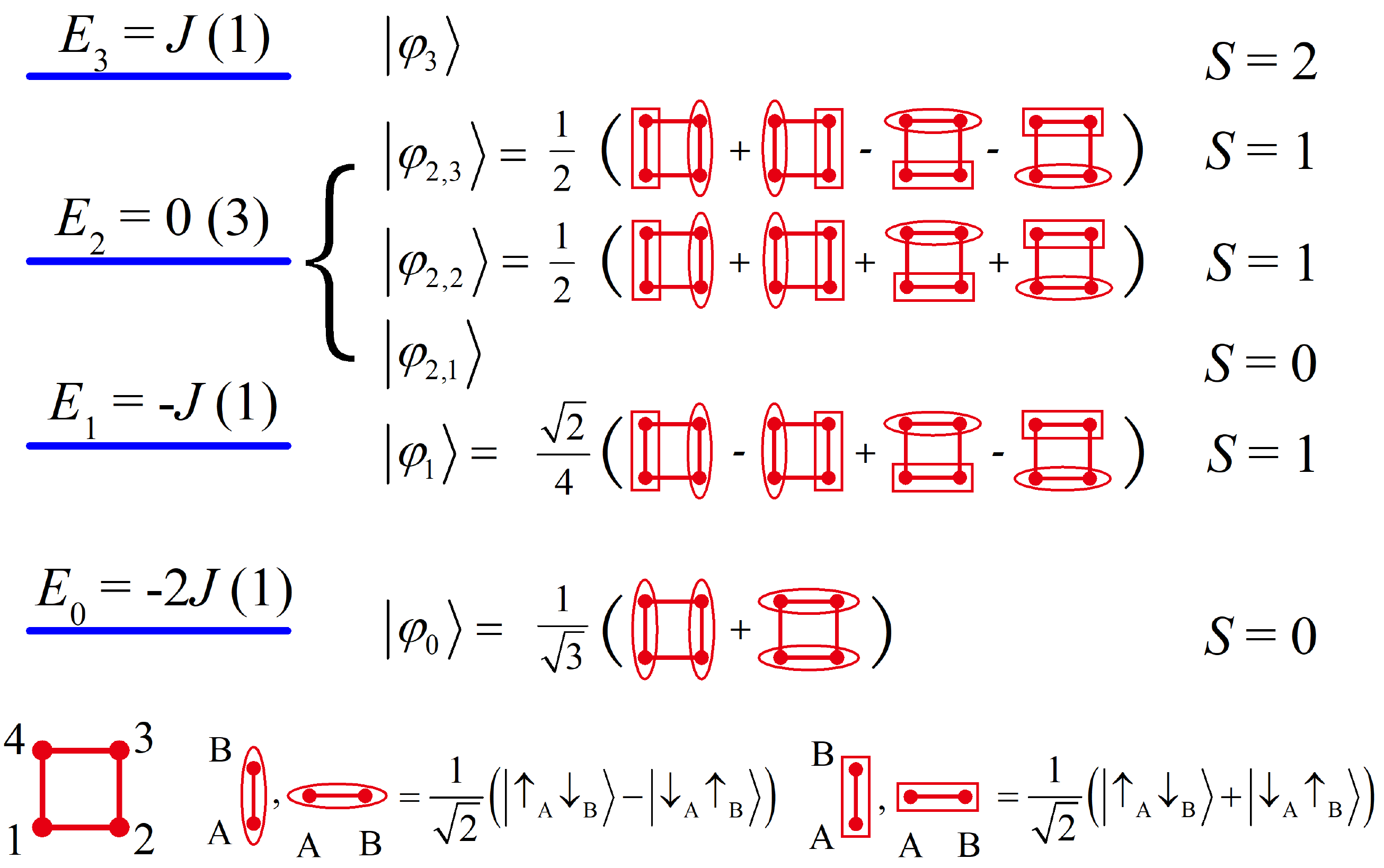}
  \caption{The energy spectrum of the Heisenberg model on a four-site plaquette in the $M_z$ = 0 sector. The corresponding spin structures of the ground state and two kinds of triplet excitations are represented by the direct product of singlets (marked by oval box) and triplets (marked by rectangular box). And we show the corresponding $S$ of different eigenstates in the rightmost of this figure, where $S$ is the magnitude of the total spin angular momentum, $\sqrt{S(S+1)}\hbar$.}
  \label{fig:App_Plaquette}
\end{figure}

The low-energy spectra on the 32- and 36-site clusters are shown in Fig.~\ref{fig:App_Energy_0p5}, which are obtained by ED using the translational and spin-inversion symmetries.
Because $N/2$ are both even for the 32- and 36-site clusters, the total spin $S$ of the states in the sectors with $z$ = 1 are even number and $S$ are odd number, on the contrary~\cite{Sandvik2010}.
In the low-lying energy spectrum, the states with $z$ = 1 are mainly singlet states, and the states with $z$ = -1 are mainly triplet states.
For these two clusters, the ground states are both the singlet states located in the sector with $k$ = $\Gamma$, $z$ = 1.
And the first excited states are both the singlet states located in the sector with $k$ = $X$, $z$ = 1, which is twofold degenerate.
But the lowest triplet states are located in the sectors with different $k$ for these two clusters: $k$ = (3$\pi$/4, 3$\pi$/4) which is fourfold degenerate for the 32-site cluster and $k$ = ($\pi$, $\pi$) ($M$) which is only onefold degenerate for the 36-site cluster.

\section{2$\times$2 checkerboard model}
\label{App:Checkerboard}

In Ref.~\cite{ShLYu2018}, the CPT method has been applied on the chain, ladder, and square-lattice models, in which the results successfully capture the magnon excitations in the magnetic phase as well as the deconfined spinons in the QSL phase.
In order to check the performance of the CPT method for the VBS phase, especially for the PVBS phase, we use CPT to calculate the dynamical spin structure factors $S^{+-}(\mathbf{q},\omega)$ for the 2$\times$2 checkerboard model and compare with the previous quantum Monte Carlo results~\cite{YXu2019}.
In this model, there are a N\'eel phase and a plaquette phase, and the phase transition happens at $g_c$ = 0.548524(3)~\cite{XRan2019}, where $g = J' / J$.

Figure~\ref{fig:App_Checkerboard} shows the CPT results of $S^{+-}(\mathbf{q},\omega)$ for the 2$\times$2 checkerboard model.
For a single four-site plaquette with only the nearest-neighbor interaction, there are two kinds of triplet excitations, which are shown in Fig.~\ref{fig:App_Plaquette}.
When the interplaquette interaction $J'$ is turned on and weak, the ground state is a plaquette phase and there are still two split triplet excitations in the excitation spectrum.
The upper one keeps at $\omega \approx$ 2.0 around the $X$ and ($\pi$/2, $\pi$/2) points.
The lower triplet excitation is also gapped and the excitation gaps at the $M$ and $X$ points decrease with the increasing of $g$.
At $g \approx g_c$, the ground state turns into the N\'eel phase and there is a gapless magnon excitation at the $M$ point.
Our CPT results agree well with the quantum Monte Carlo results~\cite{YXu2019}, which suggests that the CPT method can successfully capture the excitations in the PVBS phase.

\bibliography{Square-J1-J3}

\end{document}